\documentclass[a4paper,11pt]{article}
\pdfoutput=1 

\usepackage{jcappub} 
\usepackage[T1]{fontenc}
\usepackage{here}
\usepackage{siunitx}
\usepackage{ulem}

\title{\boldmath Reheating process in the \texorpdfstring{\(R^2\)}{R2} inflationary model with the baryogenesis scenario}
\date{\today}

\author[a,b]{Hyun Jeong,}
\author[a,b]{Kohei Kamada,}
\author[e,f]{Alexei A. Starobinsky,}
\author[a,b,c,d]{and Jun'ichi Yokoyama}

\affiliation[a]{\textit{Research Center for the Early Universe (RESCEU), Graduate School of Science,  The University of Tokyo, Tokyo 113-0033, Japan}}
\affiliation[b]{\textit{Department of Physics, Graduate School of Science, The University of Tokyo, Tokyo 113-0033, Japan}}
\affiliation[c]{\textit{Kavli Institute for the Physics and Mathematics of the Universe (Kavli IPMU), WPI, UTIAS, The University of Tokyo, Kashiwa, Chiba 277-8568, Japan}}
\affiliation[d]{\textit{Trans-Scale Quantum Science Institute, The University of Tokyo, Tokyo 113-0033, Japan}}
\affiliation[e]{\textit{L. D. Landau Institute for Theoretical Physics RAS, Chernogolovka, Moscow region 142432, Russian Federation}}
\affiliation[f]{\textit{Kazan Federal University, Kazan 420008, Republic of Tatarstan, Russian Federation}}

\emailAdd{jeong\_hyun@resceu.s.u-tokyo.ac.jp}
\emailAdd{kohei.kamada@resceu.s.u-tokyo.ac.jp}
\emailAdd{alstar@landau.ac.ru}
\emailAdd{yokoyama@resceu.s.u-tokyo.ac.jp}

\abstract{Post-inflationary evolution and (re)heating of the viable inflationary model, the $R^2$ one, is made more realistic by including the leptogenesis scenario into it. For this purpose, right-handed Majorana neutrinos with a large mass are added to the matter sector of the Standard Model to explain the neutrino oscillation experiments and the baryon asymmetry of the Universe. We have found parameters that characterize this model: non-minimal coupling of the Higgs field $\xi$, the mass of the right-handed Majorana neutrino $M_{N_\alpha}$ and the Yukawa coupling matrix components for the right-handed Majorana neutrino. We have analyzed the effect of these parameters on the reheating process and leptogenesis in this model and how they affect the resultant physical quantities: spectral parameters of primordial perturbations and baryon asymmetry.}

\begin{document}
\maketitle
\flushbottom

\section{Introduction}
\label{sec:introduction}

The inflationary scenario \cite{starobinsky1980new,sato1981first,guth1981inflationary,linde1982new,albrecht1982cosmology} is now a successfully established paradigm for the early Universe, which not only solves the horizon and flatness problems but also makes definite predictions for primordial cosmological perturbations that can be directly tested by observations. A lot of inflationary models have been proposed historically but observational data have already excluded many of them including once-popular models \cite{sato2015inflationary}. Still, a number of successful inflationary models including the original \(R^2\), or Starobinsky model~\cite{starobinsky1980new} remain which satisfy all presently available data. The most important and discriminatory among these predictions are the spectral index \(n_\mathrm{s}(k)\) of the power spectrum of scalar perturbations and the tensor-to-scalr ratio \(r(k)\). The generic and observationally confirmed prediction of all viable slow-roll inflationary models, that do not require consideration of post-inflationary evolution, is that 
both \(|n_\mathrm{s}(k)-1|\) and \(r(k)\) are small and weakly scale-dependent, and \(r(k)\) is of the order of \(|n_\mathrm{s}(k)-1|\) or less. But calculation of these parameters with the accuracy better than 10\% needed to match the existing and expected observational accuracy requires analyzing (re)heating process, which turns the inflationary stage into the radiation-dominated one by decay of an inflaton (dubbed the scalaron in the \(R^2\) model) into ultra-relativistic particles, anti-particles and radiation. Thus, the detailed study of the reheating process is very important for the determination of the observationally consistent theory of inflation which describes our Universe. \par
In the successful \(R^2\) inflationary model \cite{starobinsky1980new}\footnote{See also \cite{Starobinsky:1981vz} for more details including reheating and \cite{Starobinsky:1983zz} for the final primordial spectra of scalar and tensor perturbations, as well as \cite{Vilenkin:1985md,Mijic:1986iv,Maeda:1987xf} for further developments.}, the gravity sector is modified such that the \(R^2\) term is added to the Einstein-Hilbert action where \(R\) is the Ricci scalar. This was suggested long time ago by the consideration of the renormalization of an average value of the energy-momentum tensor of quantum matter fields in an external classical curved space-time, see e.g. \cite{Zeldovich:1971mw,Parker:1974qw,birrell1984quantum} and also \cite{Nariai:1971sv}. However, the smallness of the measured Fourier power spectrum of primordial scalar perturbations requires the dimensionless coefficient (\(\hbar=c=1\)) of this term to be unexpectedly very large: \(\approx 5.1\cdot 10^8\)~\cite{Starobinsky:1983zz,faulkner2007constraining,netto2016stable}\footnote{This value is given for the number of e-folds \(N=55\) of the observable part of inflation between the pivot scale \(k_0=\SI{0.05}{Mpc^{-1}}\) and the end of inflation, otherwise it has to be multiplied by \((N/55)^2\).}. On the other hand, no observable effect suggests that the coefficient in front of the Weyl tensor squared term in the effective action, also required for the one-loop renormalizability, can be so large and even may be significantly different from unity. This provides us with a possibility to use the inflationary solution of the model in the regime when the \(R^2\) term dominates the \(R\) (Einstein) one, while the undesired (due to the ghost appearance would be taken in full) Weyl squared term may be neglected at curvatures during inflation much less than the Planckian one.

If the main channel of scalaron decay after \(R^2\) inflation is to light scalar particles minimally or weakly (but not conformally) coupled to gravity, this model has a rather mild reheating process compared to other viable inflationary models such as Higgs inflation \cite{cervantes1995induced,bezrukov2008standard,Barvinsky:2008ia,bezrukov2009initial,Barvinsky:2009fy,garcia2009preheating,ema2017violent,decross2018preheating}, and the calculated spectral parameters \((n_\mathrm{s},r)\) are in agreement with the Cosmic Microwave Background (CMB) observations inside their present error bars~\cite{Planck:2018jri}.
 
In fact, the dependence of these parameters on the number of e-folds \(N\) during Higgs inflation is essentially the same as for \(R^2\) inflation, see \cite{He:2018gyf} for the explanation of this fact. It is only the actual value of \(N\) that is larger for Higgs inflation by several percents due to faster reheating after it caused by large non-minimal coupling of the Higgs boson to gravity (\(\xi\approx 4.5\cdot 10^4\sqrt \lambda\) for \(N=55\)). Furthermore, it has been argued that in the \(R^2\) inflationary model, dark matter and the right-handed Majorana neutrino can be generated by scalaron decay, so it can be made compatible with present matter content of the Universe \cite{gorbunov2011scalaron,gorbunov2012free}. \par
However, a more comprehensive study of reheating process after inflation is needed to derive the values of \(N\) and \(n_s-1\) with accuracy of about \(1\%\) or better expected in future CMB and large-scale structure observations~\cite{CMB-S4:2016ple,Abazajian:2019eic,LiteBIRD:2022cnt}. For this purpose, more possible channels of scalaron decay after inflation into particles and antiparticles of different quantum matter fields suggested by modern extensions of the Standard Model (SM) of elementary particles have to be investigated. So far, mostly the simplest channels like decay into light scalar particles minimally or non-minimally coupled to gravity have been considered. 
The enhancement of production of a certain matter field via inflaton decay means the suppression of the production of other matter fields. Thus, branching ratios of different decay channels become very important for the determination of observables such as the spectral parameters, as well as the amounts of baryon asymmetry and dark non-baryonic matter in the Universe.\par
An adjustment of the branching ratios can be done by controlling additional parameters of the inflationary model characterizing the effective coupling of inflaton (irrespective of its nature) to other particles in some extensions of SM. In the case of the \(R^2\) inflationary model with the SM matter sector, as shown in \cite{gorbunov2013r2}, the non-minimal coupling parameter of the Higgs field \(\xi\) plays the role of regulating the branching ratio between the scalaron decay into the Higgs boson and the decay into the gauge boson. When the Higgs field is conformally coupled to the Ricci scalar, the reheating is mainly realized by the scalaron decay into the gauge bosons rather than that into the Higgs boson, which leads to deviation of the reheating temperature and the spectral parameters from the well-known values. \par
In this paper, we analyze how the reheating process of the successful \(R^2\) inflationary model is modified when we add the successful leptogenesis scenario to it by introducing supermassive right-handed Majorana neutrinos \cite{fukugita1986barygenesis}. In this case, the \(R^2\) model is equipped with the SM matter fields and the three generations of the right-handed Majorana neutrinos. Thus, the masses and the Yukawa coupling matrix for the Majorana neutrinos,  \(M_{N_\alpha}\) and \(y_{\alpha \beta}\),  are the new model parameters in addition to the non-minimal coupling \(\xi\). We analyze how the reheating process and the resultant physical quantities change by varying these parameters. \par

This paper is organized as follows. In \S \ref{sec:model_description}, we introduce the \(R^2\) inflationary model and all the matter fields in an extension of the SM of elementary particles. The decay rates of scalaron into matter fields are calculated there, too. The parameters that characterize this model are the non-minimal coupling \(\xi\), the masses of the right-handed Majorana neutrinos \(M_{N_\alpha}\) and the Yukawa couplings for them \(y_{\alpha \beta}\). The latter two parameters determine the active neutrino mass. We restrict ourselves to two typical cases for the origin of the hierarchy in the active neutrino masses, case A and case B. In the case A, we assume that its hierarchy is explained by the hierarchy of the masses of the right-handed Majorana neutrino, and the Yukawa coupling matrix is arranged accordingly. In the case B, we assume that it is explained by the hierarchy in the Yukawa coupling matrix components. In the latter case, the expression for the lepton asymmetry generated by one right-handed Majorana neutrino decay, \(\delta\), becomes simple \cite{fukugita1986barygenesis,asaka1999leptogenesis}. We study the case A first. 
In \S \ref{sec:boltzmaneq}, we analyze the effect of the non-minimal coupling \(\xi\) on the reheating process by numerically solving the set of Boltzmann equations. In \S \ref{sec:mass_dependence}, we analyze the effect of the Majorana mass \(M_{N_\alpha}\) on the reheating process by solving the Boltzmann equations. We see that the mass dependence appears only when the Higgs field is conformally coupled. In \S \ref{sec:Connection_to_the_observation}, we discuss the parameter dependence of the physical quantities such as the spectral parameters and the baryon asymmetry. As a result, we find that the mass dependence strongly appears in the conformally coupled case but not so much in the minimally coupled one. In \S \ref{sec:Yukawa_matrix}, we change the configuration of the Yukawa coupling matrix of the right-handed Majorana neutrino to the case B and see how the reheating process changes with respect to the Yukawa coupling parameters \(y_{\alpha \beta}\). We found that in certain parameter spaces, the energy density of Majorana neutrino can dominate the Universe right after the conventional end of the reheating by the inflaton. In \S \ref{sec:Conclusion_Discussion}, we summarize the results obtained and discuss directions of future research. 
In this paper, 
we restrict ourselves to two typical values of \(\xi\) throughout this paper: minimal coupling (\(\xi = 0\)) and conformal coupling (\(\xi=-1/6\)).

\section{Decays during the reheating}\label{sec:model_description}

We extend the \(R^2\) inflationary model \cite{starobinsky1980new} to incorporate leptogenesis \cite{fukugita1986barygenesis} with heavy right-handed Majorana neutrinos, which induce the type-I see-saw mechanism. During the reheating epoch, there are three main candidates into which scalaron can decay: Higgs bosons, gauge bosons, and Majorana neutrinos.\footnote{The decay rates into other massless SM fermions do not have to be cared because their decay rates through the free field Lagrangian is zero in the absence of the mass term (see Eq. (\ref{eq:decay_rates_majorna})) and the decay channel through the Yukawa interaction is phase space suppressed. } In this section, we will see that the model parameters that govern the reheating process are the non-minimal coupling \(\xi\) and the mass of the right-handed Majorana fermion \(M_{N_\alpha}\). The Yukawa coupling matrix of the right-handed Majorana neutrino \(y_{\alpha \beta}\) 
is also important for the completion of reheating, but we take the assumption of case A for the Yukawa coupling matrix so that its parameters does not appear explicitly.

\subsection{Decays of the scalaron}\label{subsec:decays_of_scalaron}

We consider the following action defined in the Jordan frame,

\begin{equation}
    S^\text{JF} = S_{\text{Gravity}}^{\text{JF}}+S_{\text{SM}}^{\text{JF}}+S_{\text{Majorana}}^{\text{JF}},   
\end{equation}
where
\begin{equation}
    \label{eq:gravity_JF}
    S_{\text{Gravity}}^{\text{JF}}=\frac{M_{\text{G}}^2}{2} \int d^4x \sqrt{-g}  \left(R+\frac{R^2}{6M^2}\right),
\end{equation}

\begin{equation}
    \label{eq:Higgs_JF}
    \begin{split}
        S_{\text{SM}}^{\text{JF}} 
        &= \int d^4x \sqrt{-g}
        \left[- g^{\mu\nu} D_\mu \mathcal{H}^\dagger  D_\nu \mathcal{H} 
        - m_h^2 \mathcal{H}^\dagger \mathcal{H} \right. \\
        &\left.- \lambda \left(\mathcal{H}^\dagger \mathcal{H}\right)^2
        +\xi R\mathcal{H}^\dagger \mathcal{H} -\frac{1}{4} \sum_{F} F^a_{\mu\nu} F^{a\mu \nu} +\cdots \right],
    \end{split}
\end{equation}
and
\begin{equation}
    \label{eq:majorana_JF}
    \begin{split}
        S_{\text{Majorana}}^{\text{JF}} 
        &= \sum_{\alpha} \int d^4x \sqrt{-g} \left[i N^{\dagger}_\alpha \bar{\sigma}^{\mu} \nabla_\mu N_\alpha -\left( \left(\frac{1}{2} M_{N_\alpha} N_\alpha N_\alpha
        +y_{\alpha \beta} N_\alpha \tilde{\cal H}^\dagger L_\beta \right) +\text{h.c.} \right) \right].
    \end{split}
\end{equation}
Here, \(\mathcal{H} \), \(N_\alpha\) ,  and \(L_\beta\) are the SU(2) doublet Higgs field, the right-handed Majorana neutrino field (two-component spinor) and left-handed lepton doublets, respectively. In (\ref{eq:gravity_JF}),  \(M_{\text{G}}=\SI{2.4e18}{GeV}\) is the reduced Planck mass and \(M=1.3\times 10^{-5} M_{\text{G}}=\SI{3.1e13}{GeV}\) is the mass of the scalaron, which is unambiguously related to the magnitude of the primordial power spectrum of scalar perturbations determined from the measured CMB temperature and polarization fluctuations  \cite{faulkner2007constraining}. In (\ref{eq:Higgs_JF}), \(D_\mu\) is the general coordinate covariant and gauge covariant derivative, and \(\sum_{F}\) means the sum over U(1)\(_Y\), SU(2)\(_W\) and  SU(3)\(_c\)  gauge fields in SM. The non-minimal coupling of the Higgs field to Ricci scalar \(\xi R \mathcal{H}^\dagger \mathcal{H}\), which is necessary for the renormalization of the ultraviolet divergence of the quantum Higgs field on the classical curved space-time, is introduced. 
In (\ref{eq:majorana_JF}), \(\tilde{\mathcal{H}} = \epsilon_{\alpha \beta} {\cal H}^{*\beta}\). \(M_{N_\alpha}\) and \(y_{\alpha \beta}\) are the mass and components of the Yukawa coupling matrix, respectively. In the matrix \(y_{\alpha \beta}\), the row index \(\alpha\) runs over the three generations of the Majorana neutrino and the column index \(\beta\) runs over the three generations of the left-handed lepton doublets. 
\(\nabla_\mu\) is the covariant derivative defined in the curved space, which includes the spin connection \cite{birrell1984quantum}. \(\bar{\sigma}^\mu = \left(\boldsymbol{1}, -\boldsymbol{\sigma}\right)\) is the four-vector Pauli matrices.  \par

It can be easily seen that this model has a mechanism of inflation by performing
the following redefinition of the metric tensor field\footnote{We can see this fact even in the Jordan frame when we examine the classical equation of motion \cite{starobinsky1980new,Starobinsky:1981vz}. }:

\begin{equation}
    g_{\mu\nu}\to \tilde{g}_{\mu\nu} = \Omega^2 g_{\mu\nu} = \exp\left(\sqrt{\frac{2}{3}} \frac{\phi}{M_{\text{G}}}\right) g_{\mu\nu}.
\end{equation}
Then, those actions are transformed into the Einstein frame \cite{Maeda:1987xf,maeda1989towards}, and the new degree of freedom \(\phi\) (scalaron) can be extracted from the metric (in addition to two gravitons) when the action has the form of (\ref{eq:gravity_JF}):

\begin{equation}
        S_{\text{Gravity}}^{\text{EF}} 
        = \int d^4x \sqrt{-\tilde{g}}\\
        \left[\frac{M_{\text{G}}^2}{2}\tilde{R} -\frac{1}{2} \tilde{g}^{\mu\nu} \partial_\mu \phi \partial_\nu \phi -V (\phi) \right] .
\end{equation}
The potential \(V (\phi) \) is calculated as 

\begin{equation}
    V (\phi) = \frac{3M^2 M_{\text{G}}^2}{4}  \left[ 1- \exp \left( -\sqrt{\frac{2}{3} }  \frac{\phi}{M_{\text{G}}}\right)\right]^2 .
\end{equation} 
Consequently, slow-roll inflation can take place for large field values of \(\phi\), where the potential is sufficiently flat. After the end of inflation, the scalaron starts to oscillate around the origin of the potential and begins to decay into matter fields. The decay channels can be found from the relevant part of the action expanded up to the linear order in \(\phi\) in the Einstein frame\footnote{Strictly speaking, this is not the Einstein frame because we have the term \(\xi \tilde{R} \tilde{\mathcal{H}}^\dagger \tilde{\mathcal{H}}\) other than the Einstein-Hilbert term. Nevertheless, we call this frame Einstein frame just for convenience. } : 
\begin{equation}
    \label{eq:Higgs_EF}
    \begin{split}
        S_{\text{SM}}^{\text{EF}} 
        &\simeq \int d^4x \sqrt{-\tilde{g}} \left[ -\tilde{g}^{\mu\nu} \partial_\mu \tilde{\mathcal{H}}^\dagger \partial_\nu \tilde{\mathcal{H}} 
        -m_h^2 \tilde{\mathcal{H}}^\dagger \tilde{\mathcal{H}} 
        +\frac{2}{\sqrt{6}} m_h^2 \tilde{\mathcal{H}}^\dagger \tilde{\mathcal{H}} \frac{\phi}{M_{\text{G}}} \right. \\
        &+\left.\xi \tilde{R} \tilde{\mathcal{H}}^\dagger \tilde{\mathcal{H}}
        - \lambda \left(\tilde{\mathcal{H}}^\dagger \tilde{\mathcal{H}}\right)^2
        - 2\sqrt{6} \left(\xi+\frac{1}{6} \right) 
        \tilde{g}^{\mu\nu} \frac{\partial_\mu \phi}{M_{\text{G}}} \tilde{\mathcal{H}} \partial_\nu \tilde{\mathcal{H}} \right. \\
        &-\left.  \left(\xi+\frac{1}{6}\right)
        \tilde{g}^{\mu\nu} \frac{\partial_\mu \phi \partial_\nu \phi}{M_{\text{G}}^2} \tilde{\mathcal{H}}^\dagger \tilde{\mathcal{H}}  +\cdots \right],
    \end{split}
\end{equation}

\begin{equation}
    \label{eq:Majorana_EF}
    \begin{split}
        S_{\text{Majorana}}^{\text{EF}} 
        &\simeq \sum_{\alpha} \int d^4x \sqrt{-\tilde{g}} \left[i \tilde{N}^\dagger_\alpha \tilde{\bar{\sigma}}^\mu \tilde{\nabla}_{\mu} \tilde{N}_\alpha 
        -\frac{1}{2} M_{N_\alpha} \left(\tilde{N}_\alpha \tilde{N}_\alpha +\tilde{N}^\dagger_\alpha \tilde{N}^\dagger_\alpha \right) \right.\\
        &+\left.\frac{1}{2\sqrt{6}} \frac{M_{N_\alpha}}{M_{\text{G}}} \phi \left(\tilde{N}_\alpha \tilde{N}_\alpha+ \tilde{N}^\dagger_\alpha \tilde{N}^\dagger_\alpha \right)+\cdots \right].
    \end{split}
\end{equation}
Fields with a tilde are the Weyl-transformed fields\footnote{In this paper, the Weyl transformation refers to the scaling of the metric tensor field, on the other hand, the conformal transformation is the coordinate transformation which induces the Weyl scaling of the metric.} defined by \(\tilde{\mathcal{H}} = \Omega^{-1} \mathcal{H}\) in (\ref{eq:Higgs_EF}) and \(\tilde{N_{\alpha}} = \Omega^{-3/2} N_{\alpha}\) in (\ref{eq:Majorana_EF}). 
In (\ref{eq:Higgs_EF}), there are terms that were absent in the Jordan frame (\ref{eq:Higgs_JF}) because, in general, the matter action is not Weyl invariant. Indeed, we generally have the coupling between the scalaron and the trace part of the energy-momentum tensor as follows to the lowest order in \(\phi\):
\begin{equation}
\label{eq:trace}
    S_{\text{matter}}^{\text{EF}} =  \frac{1}{\sqrt{6}}\int d^4x \sqrt{-\tilde{g}} \frac{\phi}{M_{\text{G}}} T^\mu_\mu + O(\phi^2),
\end{equation}
where the terms in (\ref{eq:Higgs_EF}) and (\ref{eq:Majorana_EF}) are part of it. Note that the trace of the energy-momentum tensor reflects the conformally-non-invariant part of the model.

\par
There are also quantum contributions to the trace part of the energy-momentum tensor (\ref{eq:trace}), which come from the fact that there are no regularization schemes that hold conformal invariance simultaneously. Using the dimensional regularization, the quantum contribution, or anomaly is calculated as \cite{watanabe2011rate,gorbunov2013r2,morozov1986anomalies,Kamada:2019pmx}:

\begin{equation}
\label{eq:trace_anom}
\begin{split}
    &\left(T^\mu_\mu \right)_{\text{quantum}}
    = \epsilon \left(-\frac{1}{4} \sum_{\tilde{F}} \tilde{F}^{a}_{\mu \nu} \tilde{F}^{a\mu\nu} -\lambda \left(\tilde{\mathcal{H}}^\dagger \tilde{\mathcal{H}}\right)^2 \right)\\
    &\xrightarrow{\epsilon\to 0} 
    - \sum_{\tilde{F}} \frac{\beta (\alpha_{\tilde{F}})}{32\pi^2} \tilde{F}^{a}_{\mu \nu} \tilde{F}^{a\mu\nu} - \beta (\lambda) \left( \tilde{\mathcal{H}}^\dagger \tilde{\mathcal{H}}\right)^2 ,
\end{split}
\end{equation}
where $\beta (\alpha_{\tilde{F}})$ and $\beta (\lambda)$ are the beta functions of the gauge couplings and Higgs quartic coupling, respectively. Note that \(\tilde{F}^{a}_{\mu \nu} \tilde{F}^{a\mu\nu} = F^{a}_{\mu \nu} F^{a\mu\nu}\). Based on those actions in the Einstein frame, decay rates of scalaron are calculated as follows. \footnote{There are also contributions from the running of the non-minimal coupling \(\xi\). This contributes to the anomaly of the Higgs field. However, we suppose that this contribution is negligible for the following reasons. In the minimally coupled case, the anomaly contribution is far subdominant compared to the classical contribution (\ref{eq:decay_rates_higgs}). In the case of the conformal coupling, the classical decay rate is so small that there is a possibility that the quantum contribution can come into play, but the beta function at the conformal coupling is \(0\) at the one-loop level. There are inhomogeneous terms at the higher-loop level, but this contribution is expected to be small enough to be neglected \cite{Kamada:2019hpp,Kamada:2019euz}. We thank A.~Kamada for clarifying this point.} 

\begin{equation}
    \label{eq:decay_rates_higgs}
    \Gamma_{\phi\to h} \simeq 
  \frac{M^3\sqrt{1-\frac{4m_h^2}{M^2}}}{48 \pi M_{\text{G}}^2} \left[-\left(1+6\xi\right) + \frac{2m_h^2}{M^2} \right]^2,
\end{equation}

\begin{equation}
    \label{eq:decay_rates_majorna}
    \Gamma_{\phi\to N_{\alpha}} 
    = \frac{M}{96\pi}\left(\frac{M_{N_\alpha}}{M_{\text{G}}}\right)^2
    \left(1-\frac{4M_{N_\alpha}^2}{M^2}\right)^{\frac{3}{2}},
\end{equation}

\begin{equation}
    \label{eq:decay_rates_gauge}
    \Gamma_{\phi\to g} \simeq \sum_{F} \frac{b_{\alpha_F} ^2 \alpha_F^2 \mathcal{N}_{\alpha_F}}{768\pi^3} \frac{M^3}{M_{\text{G}}^2},
\end{equation}
where the subscripts \(h, N_\alpha\) and \(g\) stand for the complex Higgs doublet, Majorana neutrinos, and gauge bosons, respectively, and \(\mathcal{N}_{\alpha_F}\) is the number of gauge bosons for corresponding interactions: \(\mathcal{N}_{\alpha_Y}=1, \ \mathcal{N}_{\alpha_W}=3, \ \mathcal{N}_{\alpha_c}=8\). \(b_{\alpha_F}\) is the first coefficient of the beta functions for the corresponding gauge fields in SM: \(b_{\alpha_Y}=41/6, \ b_{\alpha_W}=-19/6, \ b_{\alpha_c}= -7\). The decay rate (\ref{eq:decay_rates_majorna}) can be calculated by using two-component formalism \cite{dreiner2010two,case1957reformulation,pal2011dirac}, and this is half of the decay rate into four component Dirac spinor, which is due to the fact that the degree of freedom into which the scalaron can decay is decreased by half.\footnote{The formulas (\ref{eq:decay_rates_higgs}) and (\ref{eq:decay_rates_majorna}) in the limit \(m_h\to 0,~M_{N_\alpha}\to 0\) for scalaron decay into light conformally and minimally coupled scalars and fermions were present already in the early papers \cite{starobinsky1980new,Starobinsky:1981vz}.} Note that there is no decay of scalaron into two gravitons in the absence of SM interaction (\(\alpha_F=0)\)~\cite{Starobinsky:1981zc,Ema:2016hlw}. 
This decay still becomes possible due to the term proportional to the square of the Weyl tensor in the conformal trace anomaly (\ref{eq:trace_anom})~\cite{Starobinsky:1981zc}, but it can be neglected since its rate is typically suppressed by the ratio \(M^4/M_{\text{G}}^4\) compared to (\ref{eq:decay_rates_higgs}). There are also loop contributions to the \(\Gamma_{\phi\to g}\), but they can be neglected, too, because the mass of the Higgs boson is very small~\cite{Kamada:2019pmx}.  

The dots in (\ref{eq:Higgs_JF}) represent other interactions in SM such as Yukawa interactions, and these will lead to four-leg interactions in the Einstein frame (dots in (\ref{eq:Higgs_EF})). The scalaron decay through these coupling terms is phase space suppressed, so these contributions can be neglected. This suppression is the same for the decay via the second anomaly term in (\ref{eq:trace_anom}). \par

\begin{figure}[H]
    \centering
    \includegraphics[width=8.5cm,clip]{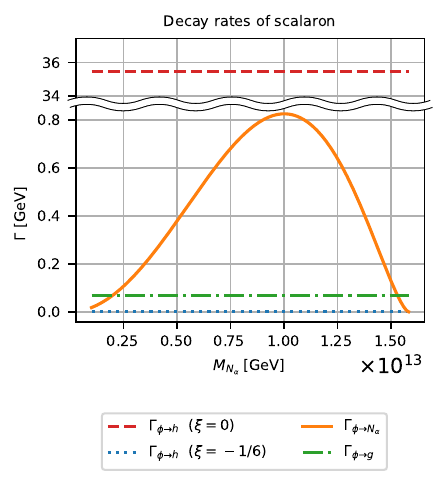}
    \caption{The decay rates of the scalaron into various fields as a function of the Majorana mass. The red dashed line and the blue dotted line represent the decay rate into the Higgs field with the minimal coupling (\(\xi =0\)) and the conformal coupling (\(\xi = -1/6\)), respectively. The orange solid line and the green dash-dot line represent the decay rate into the right-handed Majorana neutrino and the gauge fields, respectively. 
    }
    \label{fig:decayrate}
\end{figure}

The decay rates of scalaron ((\ref{eq:decay_rates_higgs}) (\ref{eq:decay_rates_majorna}), and  (\ref{eq:decay_rates_gauge})) are calculated and shown in Fig.~\ref{fig:decayrate} together with the mass dependence of the decay rate into the Majorana neutrino (\ref{eq:decay_rates_majorna}). This decay rate takes maximum when \(M_{N_\alpha}\simeq \SI{1.0e13}{GeV}\). In calculating the decay rate into the gauge bosons, the values of the gauge couplings in (\ref{eq:decay_rates_gauge}) are evaluated at the energy scale \(M/2\) using the results of \cite{Buttazzo:2013uya}.

When \(\xi=0\), the decay rate into Higgs bosons is by far the largest.  This is qualitatively because decay rates are determined by the degree of breaking of Weyl invariance, which is violently broken by the kinetic term of the Higgs scalar field.\footnote{The decay to Higgs field dominates the reheating process unless \(|\xi +1/6|< 0.007\) according to \cite{gorbunov2013r2}.}
In this subsection, the production of the matter fields was described by the decay of the scalaron in the Einstein frame, but this can also be understood as the gravitational particle production in the Jordan frame~\cite{Starobinsky:1981vz,Vilenkin:1985md}.

\subsection{Decay of the Majorana neutrino}\label{subsec:decay_of_majorana_neutrino}

The reheating process is realized not only by the decay of scalaron into gauge bosons and Higgs bosons but also through the decay of Majorana neutrinos, if produced substantially by the scalaron decay. The decay rate of the Majorana neutrino 
is given by

\begin{equation}
    \label{eq:decay_NtoR}
    \Gamma_{N_1\to R(\text{relativistic particles})} = \Gamma_{N_1\to lh} + \Gamma_{N_1\to \bar{l}h} = \frac{M_{N_1}}{8\pi} \sum_{\alpha} |y_{1\alpha}|^2.
\end{equation}
Here, only first generation of the Majorana neutrino \(N_1\) is considered because other generations \(N_{2,3}\) are not produced by the inflaton decay in our setup which will be explained shortly. These Majorana neutrinos are also assumed to explain the masses of three active neutrinos through the see-saw mechanism (Type I). The mass matrix for the active neutrinos is calculated as follows.

\begin{equation}\label{eq:active_mass_matrix}
    m_{\alpha \beta} = -y^\dagger_{\alpha \gamma} \frac{v^2}{2 M_\gamma} y_{\gamma \beta},
\end{equation}
where \(v = \SI{246}{GeV}\) is the vacuum expectation value of the Higgs field. We used the summation convention over the repeated indices here. We define \(m_1 < m_2 < m_3\) as the eigenvalues of this mass matrix (\ref{eq:active_mass_matrix}):

\begin{equation}
\label{eq:PMNS_matrix}
    \text{diag} (m_1, m_2, m_3) = U^\dagger_{i \alpha} m_{\alpha \beta} U_{\beta j}.
\end{equation}
Here, \(U_{\alpha i}\) is the Pontecorvo-Maki-Nakagawa-Sakata (PMNS) matrix, which diagonalizes the mass matrix \(m_{\alpha \beta}\). Now $i$ runs for the mass eigen states of active neutrinos. By introducing the rotated Yukawa coupling matrix \(\tilde{y}_{\alpha i} := y_{\alpha \beta} U_{\beta i}\), Eq.~(\ref{eq:PMNS_matrix}) is re-expressed as\footnote{The overall sign of the neutrino masses is irrelevant. }

\begin{equation}
\label{eq:diagonalized_mass_matrix}
    \text{diag} (m_1, m_2, m_3) =-\tilde{y}^\dagger_{i\gamma} \frac{v^2}{2M_\gamma} \tilde{y}_{\gamma j}.
\end{equation}

We make the following assumptions for simplicity.

    \paragraph{\(\spadesuit\) On the configuration for the active neutrino masses:} It is assumed that three active neutrinos have the normal mass hierarchy: 
    
    \begin{equation}
    \label{eq:value_of_active_neutrino_mass}
        m_1 \ll m_{\text{sol}}, \  m_2\simeq m_{\text{sol}} = \SI{0.009}{\eV},\  m_3\simeq m_{\text{atm}}=\SI{0.05}{\eV} 
    \end{equation}
Here, \(m_{\text{sol}}\) and \(m_{\text{atm}}\) are the values of the mass difference determined by the solar neutrino experiments and atmospheric neutrino experiments, respectively \cite{Super-Kamiokande:1998kpq,T2K:2014ghj,gonzalez2012global}.
    \paragraph{\(\spadesuit\) On the right-handed Majorana neutrino sector \(M_\alpha\), \(y_{\alpha \beta}\) (case A):} We postulate that the mass hierarchy of the active neutrinos above is explained by the mass hierarchy of three right-handed neutrinos:
    \begin{equation}
    \label{eq:inv_propto}
        m_1\propto M_3^{-1} \ll m_2\propto M_2^{-1} \ll m_3\propto M_1^{-1}
    \end{equation}
    with \(M_{N_1} \ll M_{N_2} \ll M_{N_3}\). 
    Such a feature may be realized if the main contribution to \(m_1, m_2, m_3\)  comes from terms with the anti-diagonal Yukawa coupling matrix components \(|\Tilde{y}_{13}|^2, |\Tilde{y}_{22}|^2, |\Tilde{y}_{31}|^2\), respectively (see Eq.~(\ref{eq:diagonalized_mass_matrix})), with their values being of similar order of magnitude. Furthermore, the left upper triangular components should be sufficiently small,

    \begin{equation}
    |\Tilde{y}_{11}|,|\Tilde{y}_{12}|,|\Tilde{y}_{21}| \ll |\Tilde{y}_{31}|,|\Tilde{y}_{22}|,|\Tilde{y}_{13}|,
    \end{equation}
to suppress the contribution from the lighter Majorana neutrinos.
The right lower off-diagonal components are irrelevant on (the hierarchy of) active neutrino mass unless they are too large, since their contributions are suppressed by the larger mass. To make baryogenesis efficient, we set they are of the same order of the anti-diagonal components, see \S ~\ref{subsec:baryon_asymmetry}. 
These assumptions define case A which will be adopted throughout \S \ref{sec:model_description}~-~\S \ref{sec:Connection_to_the_observation}. We consider the hierarchical case (case B) in \S \ref{sec:Yukawa_matrix}.
    \paragraph{\(\spadesuit\) Non-thermal leptogenesis:} The masses of the right-handed Majorana neutrino is so heavy that the leptogenesis occurs non-thermally. In this scenario, throughout the history of the Universe, the right-handed Majorana neutrino is never thermalized. 
    \paragraph{\(\spadesuit\) Only lightest Majorana neutrino is produced by the inflaton decay.} For simplicity, it is postulated that only the lightest \(N_1\) are produced by the decay of the scalaron during reheating. This can be achieved by the following mass configuration (see (\ref{eq:decay_rates_majorna}) and Fig.~\ref{fig:decayrate}): \(M_{N_1}<M/2<M_{N_2}\).

These assumptions lead to:
\begin{equation}
    \label{eq:neutrino_experiment}
    m_3 \sim  -\frac{|y_{13}|^2}{2M_{N_1}} v^2 \sim m_{\text{atm}} \simeq \SI{0.05}{eV}.
\end{equation}
Using this relation, (\ref{eq:decay_NtoR}) is re-expressed as 
\begin{equation}
\label{eq:decay_rate_NtoR_caseA}
    \Gamma_{N_1\to R} = \frac{M_1^2 m_{\text{atm}}}{4 \pi v^2}.
\end{equation}
Note that this expression (\ref{eq:decay_rate_NtoR_caseA}) 
is based on the assumption on the Yukawa coupling matrix, that is, case A. We will have a different expression when we change the assumptions on the Yukawa coupling matrix to the case B (see Eq.~(\ref{eq:decay_rate_NtoR_hierarchy})).

\section{The effect of the curvature coupling}\label{sec:boltzmaneq}

In the previous section, we have seen that the model parameters that govern the inflaton decay in this model are \(\xi\) and \(M_{N_1}\) when the Yukawa coupling matrix \(y_{\alpha \beta}\) has the structure of case A. Here, we analyze the effects of non-minimal coupling \(\xi\) in detail. There are two typical values of the curvature coupling \(\xi\): the minimal coupling (\(\xi = 0\)) and the conformal coupling (\(\xi= -1/6\)). In the minimally coupled case, the decay rate into the Higgs particle (\ref{eq:decay_rates_higgs}) is so strong that Higgs particles are produced mainly. On the other hand, in the conformally coupled case, the decay rate into the Higgs particles is strongly suppressed, and the scalaron decays into Majorana neutrino and the gauge bosons. \par
In this section, we numerically confirm this by solving the Boltzmann equations below while the mass of the Majorana neutrino \(M_{N_1}\) is fixed to be \(\SI{1.0e13}{GeV}\) so that the decay rate of scalaron into the Majorana neutrino is maximal (see (\ref{eq:decay_rates_majorna}) and Fig.~\ref{fig:decayrate}). 

\begin{equation}
    \frac{d\rho_\phi}{dt} = -3H \rho_\phi 
    -\Gamma_{\phi\to N_1} \rho_\phi
    - \Gamma_{\phi \to g}\rho_\phi
    -\Gamma_{\phi \to h} \rho_\phi,
\end{equation}

\begin{equation}
    \frac{d\rho_{N_1}}{dt} =-3H\rho_{N_1}+ \Gamma_{\phi\to N_1} \rho_\phi -\Gamma_{N_1\to R} \rho_{N_1},
\end{equation}

\begin{equation}
    \frac{d\rho_R}{dt} = 
    -4H\rho_R 
    +\Gamma_{\phi\to g} \rho_\phi 
    +\Gamma_{\phi\to h} \rho_\phi
    + \Gamma_{N_1\to R} \rho_{N_1},
\end{equation}

\begin{equation}
    H^2=\frac{\rho_\phi +\rho_r +\rho_{N_1}}{3M_{\text{G}}^2}.
\end{equation}
where \(\rho_\phi\), \(\rho_{N_1}\), \(\rho_R\) and \(H\) are the energy densities of the inflaton, the lightest right-handed Majorana neutrino, relativistic particles, and the Hubble parameter, respectively.

\subsection{The minimally coupled case}\label{subsec:minimally_coupled_case}

\begin{figure}[htbp]
    \centering
    \includegraphics[width=12cm,clip]{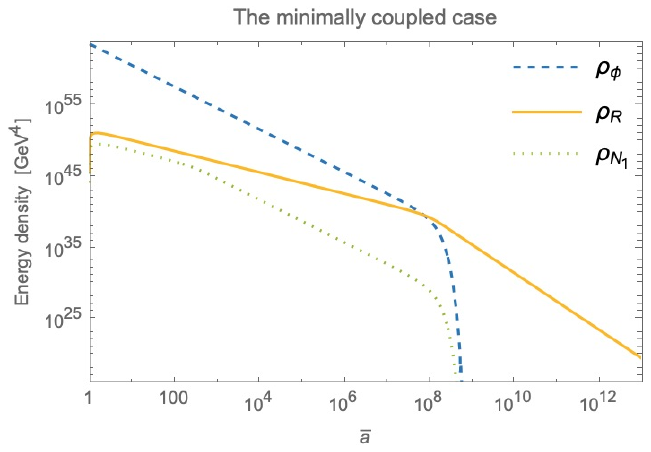}
    \caption{Evolution of the energy densities of several fields as the functions of the rescaled scale factor \(\bar{a}\) for $M_{N_1}=\SI{1.0e13}{GeV}$ and $\xi=0$. Here, we assume that the Higgs field is minimally coupled to gravity. The blue dashed line, the orange solid line and the green dotted line represent the energy densities of the inflaton, relativistic particles, and the right-handed Majorana neutrino, respectively.}
    \label{fig:boltzman_eq}
\end{figure}

For numerical calculation, the above equations are re-written with the dimensionless variables: \(\bar{a}=a/a_I, \ f = \rho_\phi M^{-4} \bar{a}^3, \ n= \rho_{N_1} M_{N_1}^{-1} M^{-3} \bar{a}^3, \ r= \rho_R M^{-4} \bar{a}^4\). 
Here, \(a_I\) is the scale factor at the end of inflation. With these variables, we can re-write the set of Boltzmann equations as follows \cite{chung1999production}.

\begin{equation}
    \frac{df}{d \bar{a}} =-\frac{ \sqrt{3} M_{\text{G}} \left(\Gamma_{\phi\to N_1}+\Gamma_{\phi\to g} +\Gamma_{\phi \to h} \right)   f  \bar{a}}{M^2 \sqrt{r+f \bar{a}+n(M_{N_1}/M)\bar{a}}} ,
\end{equation}

\begin{equation}
    \label{eq:boltzman_eq_majorana}
    \frac{dn}{d\bar{a}} = \frac{ \sqrt{3}M_{\text{G}} \left(\Gamma_{\phi\to N_1}(M/M_{N_1})f - \Gamma_{N_1\to R} n \right) \bar{a}}{M^2 \sqrt{r+f \bar{a}+n(M_{N_1}/M)\bar{a} }},
\end{equation}

\begin{equation}
\label{eq:botzman_nondim_radiation}
    \frac{dr}{d\bar{a}} = \frac{ \sqrt{3}M_{\text{G}} \left( (\Gamma_{\phi\to g}+\Gamma_{\phi\to h})f+\Gamma_{N_1 \to R} (M_{N_1}/M)n \right)\bar{a}^2}{M^2 \sqrt{r+f \bar{a} +n(M_{N_1}/M)\bar{a}}}.
\end{equation}
Once the mass of the Majorana neutrino \(M_{N_1}\) is fixed, we are ready to solve the Boltzmann equations with the initial conditions,

\begin{equation}
    f (\bar{a}=1) =3 H_{\text{inf}}^2 M_{\text{G}}^2 M^{-4}\ \text{and} \   \ n(1)=r(1)=0.
\end{equation}
where \(H_{\text{inf}}=M/2\) is the Hubble parameter at the end of inflation. The result of numerical calculation of Boltzmann equations when \(M_{N_1}\) is fixed to be \(\SI{1.0e13}{GeV}\) is shown in Fig.~\ref{fig:boltzman_eq}.

\begin{figure}[htbp]
    \centering
    \includegraphics[width=12cm,clip]{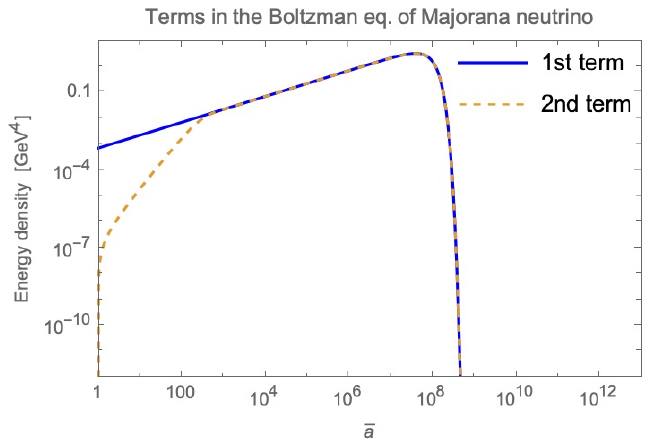}
    \caption{The decay or source contribution to the energy density of Majorana neutrino is shown with respect to \(\bar{a}\). The blue solid line and the orange dashed line represent the first term and the second term in the RHS of (\ref{eq:boltzman_eq_majorana}).}
    \label{fig:comparison}
\end{figure}

During the inflaton oscillation-dominated epoch, \(f\) remains approximately constant because the decay products' energy density is negligible compared to that of the inflaton. Thus, \(\rho_{\phi}\propto \bar{a}^{-3}\). In  this epoch, the energy density of relativistic particles is \cite{Kolb:1990vq}
\begin{equation}
    \rho_R \simeq \frac{4}{5} \Gamma_{\phi \to h} M^3 \bar{a}^{-\frac{3}{2}} .
\end{equation}
As for the Majorana neutrino \(N_1\), (\ref{eq:boltzman_eq_majorana}) can be solved as
\begin{equation}\label{eq:Majorana_analytic}
\begin{split}
        &n = \frac{3\Gamma_{\phi \to N_1} M_{\text{G}}^2}{4 \Gamma_{N_1\to R} M_{N_1} M} \left[1-\exp \left[ -\frac{4 \Gamma_{N_1\to R}}{3 M} \left(\bar{a}^{\frac{3}{2}}-1\right) \right]\right].
\end{split}
\end{equation}
When \(n\) is small and the decay term in RHS of (\ref{eq:boltzman_eq_majorana}) is negligible, the energy density of the Majorana neutrino can be approximated as
\begin{equation}
    \rho_{N_1} \simeq \Gamma_{\phi \to N_1} \left(\frac{M}{M_{N_1}}\right) M_{\text{G}}^2 M_{N_1} \bar{a}^{-\frac{3}{2}} .
\end{equation}
Gradually, the effect of the decay of Majorana neutrino into relativistic particles cannot be ignored after some growth of its energy density \(n\). After \(\bar{a}_t \simeq \SI{8.8e2}{}\), the source and decay terms in (\ref{eq:boltzman_eq_majorana}) are balanced, so  the rescaled energy density \(n\) becomes constant and \(\rho_{N_1} \propto \bar{a}^{-3}\). The comparison of the first and second terms in RHS of (\ref{eq:boltzman_eq_majorana}) are shown in Fig.~\ref{fig:comparison}. It can be seen that the matching period \(\bar{a}_t \simeq \SI{8.8e2}{}\) is consistent with the moment at which the power-law of the energy density of the Majorana neutrino changes in Fig.~\ref{fig:boltzman_eq}. 
This transition period \(\bar{a}_t\) is actually a function of the mass of the Majorana fermion \(M_{N_1}\). It depends on the mass of the Majorana fermion as
\begin{equation}\label{eq:Majorana_power_law_transition}
    \bar{a}_t=\left[-\frac{3}{2} \left(\frac{2}{3} + \frac{M}{2\Gamma_{N_1\to R}} \ln \left[\frac{3\Gamma_{\phi \to N_1} M_{\text{G}}^2}{2M_{N_1} M^2}\right]\right)\right]^{\frac{2}{3}} 
\end{equation}
which is obtained by substituting (\ref{eq:Majorana_analytic}) into \(d n /d\bar{a} = 0\) and solving for \(\bar{a}_t\). This power-law transition period (\ref{eq:Majorana_power_law_transition}) is a decreasing function of the mass of the Majorana fermion. This is reasonable because the greater the mass \(M_{N_1}\), the larger the decay rate of the Majorana neutrino \(\Gamma_{N_1\to R}\) (see Eq.~(\ref{eq:decay_rate_NtoR_caseA})) and the faster it takes to decrease by the decay of the Majorana neutrino.

As the energy density of relativistic particles \(\rho_R\) decreases slower than that of inflaton \(\rho_\phi\), there is a moment when the former reaches the latter: \(\rho_R = \rho_\phi\). From this moment, the Universe becomes radiation-dominated, which leads to the different behavior of the scale factor. We define the reheating temperature \(T_R\) as the temperature at this transition moment. 
Note that the SM interaction is rapid enough such that the relativistic particles are already thermalized in the parameters of our interest~\cite{davidson2000thermalisation,allahverdi2010reheating,Harigaya:2013vwa,Mukaida:2015ria,Mukaida:2022bbo}.
Using Stefan-Boltzman's law, \(\rho_R= (\pi^2 g_r/30) T^4\), the reheating temperature is calculated as \(T_R=\SI{3.26e9}{GeV}\) by taking into account the degrees of freedom of relativistic particles
at the reheating: \(g_r = 106.75\) (for the SM). 
Similarly, the reheating temperature in the original \(R^2\) inflation model without heavy Majorana neutrinos is calculated as \(T_R= \SI{3.22e9}{GeV}\). The enhancement in our model is reasonable because the scalaron gets two additional channels to radiation and the reheating becomes slightly more efficient. This enhancement of order of one percent in the reheating temperature can also be checked by \(\sqrt{\left(\Gamma_{\phi \to h}+\Gamma_{\phi \to N_1}+\Gamma_{\phi\to g}\right)/\Gamma_{\phi \to h}}\simeq 1.01\), in which we used the fact that the reheating temperature is proportional to the root of the total decay rate of the scalaron.
Because of the low reheating temperature \(T_R \ll M_{N_1}\), the right-handed neutrinos are not thermalized, and hence leptogenesis occurs non-thermally.

\subsection{The conformally coupled case}\label{subsec:conformally_coupled_case}

The second natural choice for the curvature coupling parameter \(\xi\) is conformal coupling \(\xi=-1/6\)\footnote{It is well-motivated in the context of supergravity, too.}, which preserves the Weyl invariance of the Higgs kinetic term in this case. Thus, we can suppress the decay of the scalaron to the Higgs particles\footnote{The decay rate through the violation of conformal invariance by the Higgs mass term is negligible.} (see Eq.~(\ref{eq:decay_rates_higgs})) and the branching ratio of the decay into the Majorana neutrino is enhanced. Therefore, in this case, the decay rate \(\Gamma_{\phi\to h}\) is negligible. \par
\begin{figure}[htbp]
    \centering
    \includegraphics[width=12cm,clip]{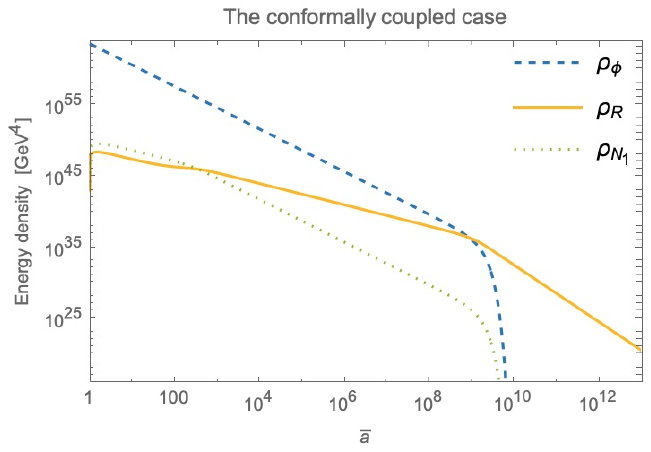}
    \caption{Results of numerical calculation of Boltzmann equations when the Higgs field is conformally coupled to \(R\) with $M_{N_1} = \SI{1.0e13}{GeV}$. The blue dashed line, the orange solid line and the green dotted line represent the energy density of the inflaton, the radiation and the Majorana neutrino, respectively.}
    \label{fig:boltzman_eq_conformal}
\end{figure}

Following the same procedure discussed in \ref{subsec:minimally_coupled_case}, the result is shown in Fig.~\ref{fig:boltzman_eq_conformal}, again fixing the mass to be \(M_{N_1} = \SI{1.0e13}{GeV}\), and assuming \(M_{N_2}, M_{N_3} > M/2\).

From Fig.~\ref{fig:boltzman_eq_conformal}, it can be seen that \(\rho_R\) is more suppressed in the conformally coupled case than in the minimally coupled case (Fig.~\ref{fig:boltzman_eq}). As a result, the transition period into the radiation-dominated era becomes later compared to the minimally coupled case. The corresponding reheating temperature is calculated as \(T_R = \SI{5.11e8}{GeV}\)\footnote{If we take the mass of the Majorana field \(M_{N_1}=\SI{3.0e12}{GeV}\), or \(\SI{1.5e13}{GeV}\) so that decay rates (\ref{eq:decay_rates_majorna}) and (\ref{eq:decay_rates_gauge}) coincide with each other, the corresponding reheating temperature is \(T_R = \SI{2.98e8}{GeV}\). This decrease is reasonable because, in the limit \(\Gamma_{\phi \to N_1}\to 0\), the reheating temperature must decrease to \(\SI{1.4e8}{GeV}\) according to \cite{gorbunov2013r2}, in which the right-handed Majorana neutrino is absent.}. In this model, the Universe is reheated mainly by the subsequent decay of the produced Majorana neutrino into radiation, unlike the minimally coupled case.

\section{The effect of the Majorana fermion mass}\label{sec:mass_dependence}
So far, we have discussed the reheating process by fixing the mass of the Majorana neutrino. In this section, we discuss the dependence on the mass of the Majorana neutrino on the reheating process. 
As we have seen, in the minimally coupled case, the reheating is governed mainly by the Higgs particles, thus the variation in the \(M_{N_1}\) changes the reheating temperature very little.\par
\begin{figure}[htbp]
    \centering
    \includegraphics[width=12cm,clip]{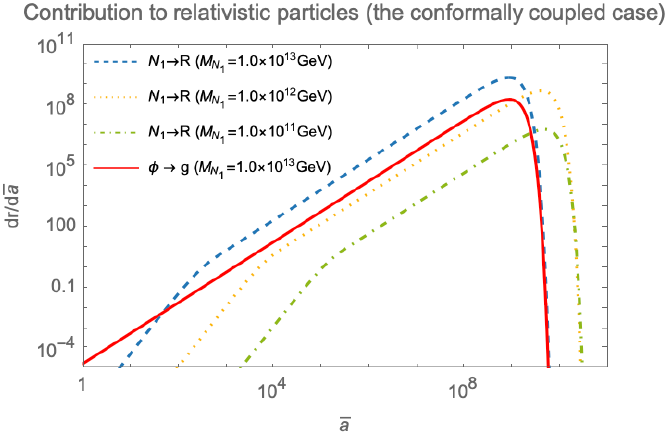}
    \caption{The contribution to radiation from each decay mode (see Eq.~(\ref{eq:botzman_nondim_radiation})) is calculated. The blue dashed line, the orange dotted line and the green dash-dot line represent the contribution from the decay of the Majorana neutrino \(N_1\) with mass \(M_{N_1} = \SI{1.0e13}{GeV}\), \(M_{N_1} = \SI{1.0e12}{GeV}\) and \(M_{N_1} = \SI{1.0e11}{GeV}\), respectively. The red solid line represent the contribution from the inflaton decay into the gauge bosons when the Majorana mass is \(M_{N_1} = \SI{1.0e13}{GeV}\). As for the contribution from the gauge bosons, the change in the Majorana mass only results in the change in the moment of the transition to the exponential decay. }
    \label{fig:contribution}
\end{figure}

On the other hand, in the conformally coupled case, the reheating is realized mainly by the production and the decay of the Majorana neutrino, thus the Majorana mass dependence appears.  We compare terms in the RHS of (\ref{eq:botzman_nondim_radiation}) below and  can see that the contribution from the decay of the Majorana neutrino is the largest if \(M_{N_1} = \SI{1.0e13}{GeV}\) as seen in Fig.~\ref{fig:contribution}.  If we decrease the mass of the Majorana fermion, the contribution from the gauge bosons exceeds that from the decay of the Majorana neutrino when the Majorana mass \(M_{N_1}\) is below \(M_{N_1} \simeq \SI{1.0e12}{GeV}\), which corresponds to the fact that the decay rate into the Majorana neutrino becomes far sub-dominant: \(\Gamma_{\phi \to N_1} \ll \Gamma_{\phi \to g}\). 

\begin{figure}[htbp]
    \centering
    \includegraphics[width=12cm,clip]{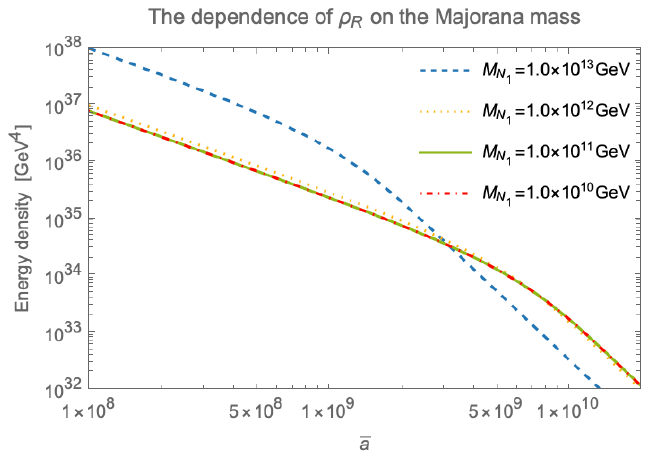}
    \caption{The blue dashed line, the orange dotted line, the green solid line and the red dash-dot line represent the evolution of the energy density of radiation \(\rho_R\) with different masses: \(M_{N_1}= \SI{1.0e13}{GeV}\), \(M_{N_1}= \SI{1.0e12}{GeV}\), \(M_{N_1}= \SI{1.0e11}{GeV}\) and \(M_{N_1}= \SI{1.0e10}{GeV}\), respectively. }
    \label{fig:mass_dependence}
\end{figure}

The consequence of this can be seen in the behavior of \(\rho_R\) on \(M_{N_1}\) near the transition of power law. We show this behavior in Fig.~\ref{fig:mass_dependence}. This mass dependence cannot be found in the minimally coupled case. \par
As we lower the mass of the Majorana neutrino \(M_{N_1}\), the energy density at the transition to radiation domination becomes smaller, resulting in the lower reheating temperature. This leads to the difference in the spectral parameters, which will be discussed in \ref{subsec:spectral_indices}. 

However, it can be seen that the mass dependence disappears below around \(\SI{1.0e12}{GeV}\) because the decay rate of Majorana fermion into relativistic particles becomes so small (see Fig.~\ref{fig:contribution}) that the reheating is switched to be governed by the scalaron decay into gauge bosons rather than by the decay of the Majorana fermion.  In this case, the only relevant decay channel is the decay into gauge bosons as we have suppressed the decay channel into Higgs bosons and Majorana fermions by taking conformal coupling and taking the small mass, respectively. \par

This mass dependence of the transition period results in the mass dependence of the reheating temperature \(T_R\). This is approximately described as \footnote{This can be obtained by naively taking the equality \(\Gamma_{\phi} = H\).}

\begin{equation}
    T_R \simeq 1.1\times 10^{9} \times \left(\frac{M_{N_1}}{\SI{e13}{GeV}}\right) \left(1-\frac{4 M_{N_1}^2}{M^2}\right)^{\frac{3}{4}} \SI{}{GeV}
\end{equation}
when \(\SI{2.0e12}{GeV} < M_{N_1}<\SI{1.5e13}{GeV} \) and the decay rate \(\Gamma_{\phi \to N_1}\) exceeds  \(\Gamma_{\phi \to g}\) (see Fig.~\ref{fig:decayrate}). If this condition is not satisfied, the mass dependence of the reheating temperature disappears to yield \(T_R \simeq \SI{2.2e8}{GeV}\). This approximated expression almost coincides with the numerical results, but we will use the numerical results for the accurate estimation of the spectral parameters.

\section{Parameter dependence of the obsevables}\label{sec:Connection_to_the_observation}

We have seen that the reheating process heavily depends on the non-minimal coupling \(\xi\). The dependence on \(M_{N_1}\) is negligiblly small in the minimally coupled case, because the scalaron mainly decays into the Higgs particles, and the branching ratio into the Majorana neutrino becomes very small. On the contrary, the mass dependence does appear in the conformally coupled case for \(M_{N_1}>\SI{1.0e12}{GeV}\) because the decay into the Higgs particles is suppressed and the branching ratio into the Majorana fermion becomes large. In the previous section, we have confirmed this by numerically solving the Boltzmann equations.
In this section, we discuss the above argument by calculating the parameter dependence of the observables, namely the spectral parameters and the baryon-to-entropy ratio.

\subsection{Spectral parameters}\label{subsec:spectral_indices}

The number of e-folds of inflation, $N_{0.002}$, corresponding to the CMB pivot scale \(k_{\text{CMB}} = \SI{0.002}{Mpc^{-1}}\) depends on the duration of the field oscillation period before 
completion of the reheating \cite{Planck:2018jri,baumann2022cosmology,sato2015inflationary} and is related to the reheating temperature as 
\begin{equation}\label{eq:efolding_number}
    N_{0.002} = N_{0.05}+\log 25 \simeq57-\frac{1}{3} \log \frac{\SI{e13}{GeV}}{T_R}.
\end{equation}
When we derive (\ref{eq:efolding_number}), we adopted an approximation that the change of the equation of state is instantaneous between the end of inflaton-dominated era and the onset of radiation-dominated era, which can be confirmed numerically. 
It can also be expressed by the integral of inflaton field value \(\phi\) as

\begin{equation}
    N_{0.002}= \frac{1}{M_{\text{G}}^2}\int_{\phi_e}^{\phi_*} d\phi \frac{V[\phi]}{V'[\phi]}, 
\end{equation}
where $\phi_*$ and $\phi_e$ are the inflaton field values when the pivot scale \(k_{\text{CMB}} = \SI{0.002}{Mpc^{-1}}\) exited the horizon and at the end of inflation, respectively.

\begin{figure}[H]
    \centering
    \includegraphics[width=11cm,clip]{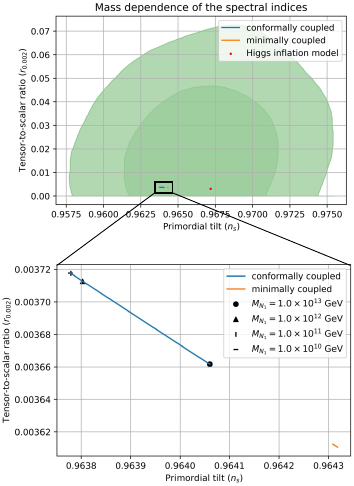}
    \caption{The blue solid line and the orange solid line represent the mass dependence of the spectral parameters \((n_s, r)\) evaluated at the pivot scale \(k_{\text{CMB}} = \SI{0.002}{Mpc^{-1}}\) in the conformally coupled case and the minimally coupled case, respectively. In the upper graph, we simultaneously plot the \(68\%\) and \(95\%\) CL regions for the spectral parameters from TT, TE, EE\(+\)lowE\(+\)lensing\(+\)BK15\(+\)BAO data \cite{Planck:2018jri}.  For comparison, we also plot the red point representing the value in the case of the Higgs inflation. In the lower graph, for the conformally coupled case, we re-plotted the value of \((n_s, r)\) when we take different values for the Majorana mass: \(M_{N_1}=\SI{1.0e13}{GeV}\), \(M_{N_1}=\SI{1.0e12}{GeV}\), \(M_{N_1}=\SI{1.0e11}{GeV}\), and \(M_{N_1}=\SI{1.0e10}{GeV}\) corresponding to the circle point, the triangle point, the vertical line and the horizontal line, respectively. }
    \label{fig:nsr}
\end{figure}

Because the curvature perturbation is conserved after the horizon crossing, the slow-roll parameters for the evaluation of the properties of the cosmological perturbation measured by the CMB are calculated by this field value \(\phi_*\) as

\begin{equation}
    \left\{
    \begin{array}{ll}
        \epsilon &= \left. \frac{1}{2}  M_{\text{G}}^2 \left(\frac{V'(\phi)}{V(\phi)}\right) \right|_{\phi = \phi_*} ,\\
        \eta &= \left. M_{\text{G}}^2 \left(\frac{V''(\phi)}{V(\phi)}\right) \right|_{\phi = \phi_*} ,\\
        \zeta^2 &= \left. M_{\text{G}}^4 \left(\frac{V'(\phi) V'''(\phi)}{V(\phi)^2}\right) \right|_{\phi = \phi_*}.
    \end{array}
    \right.
\end{equation}
The spectral parameters are calculated by using these slow roll parameters as

\begin{equation}
    \left\{
    \begin{array}{ll}
        n_s &= 1+2\eta -6\epsilon -\frac{2}{3} \eta^2 +0.374 \zeta^2, \\
        n_T &= -2\epsilon, \\
        r &= 16\epsilon.
    \end{array}
    \right.
\end{equation}

We show the mass dependence of the spectral parameters in the minimally coupled case and the conformally coupled case in Fig.~\ref{fig:nsr}. In addition to these, we show the values of the Higgs inflation model for comparison\footnote{Here we assume that the energy of the inflaton is immediately converted to the energy of the radiation after the end of inflation due to the violent preheating \cite{ema2017violent,decross2018preheating}.}.
We can see that the mass dependence disappears in the minimally coupled case. On the other hand, the mass dependence is manifest in the conformally coupled case at $M_{M_1} \gtrsim 10^{11} \mathrm{GeV}$.   As we diminish the mass of the Majorana fermion, the mass dependence disappears below around \(M_{N_1}\lesssim \SI{e11}{GeV}\). This is because gradually the main role of the reheating switches to the decay into gauge bosons as we have seen in \S \ref{sec:mass_dependence}. When $M_{M_1} \gtrsim 10^{11} \mathrm{GeV}$, in order to distinguish the right-handed neutrino mass with the future CMB observation, we require the accuracy of \(\mathcal{O}(10^{-4})\) for \(n_s\) and of \(\mathcal{O}(10^{-5})\) for \(r\).

\subsection{Baryon asymmetry}\label{subsec:baryon_asymmetry}

Since in our setup the right-handed neutrinos are copiously produced, 
the baryon asymmetry of the Universe may be generated through 
leptogenesis, which is one of the most promising baryogenesis scenarios \cite{fukugita1986barygenesis}. In this scenario, 
lepton asymmetry, which is produced when the heavy Majorana neutrino decays, is converted to the baryon asymmetry by the sphaleron process~\cite{Manton:1983nd,klinkhamer1984saddle,arnold1987sphalerons,Kuzmin:1985mm}. 
In this subsection, we investigate how the successful leptogenesis can take place.

The net lepton asymmetry produced by one \(N_1\) particle decay is expressed as: 
\begin{equation}
    \delta \equiv \frac{\Gamma (N_1\to lh) - \Gamma (N_1\to \bar{l}h)}{\Gamma (N_1\to lh) + \Gamma (N_1\to \bar{l}h)}.
\end{equation}
This asymmetry comes from the interference terms between the tree-level diagram and the one-loop diagram. 
Including both one-loop vertex and self-energy correction, we obtain~\cite{fukugita1986barygenesis,covi1996cp,asaka1999leptogenesis} 

\begin{equation}
\begin{split}
    \label{eq:microscopic_lepton_asymmetry}
    \delta 
    &\simeq -\frac{3M_1}{16\pi  y_{1\rho}y_{\rho 1}^{\dagger}}\text{Im}\left[y_{1\alpha} y_{1\beta} \left(y_{\gamma \alpha}^* \frac{1}{M_{N_\gamma}} y_{\gamma\beta}^*\right)\right] \\
    &\equiv -\frac{3\delta_{\text{eff}}}{16 \pi y_{1\rho}y_{\rho 1}^{\dagger} }  \left|\left(y_{1\alpha} y_{\alpha 2}^{\dagger}\right)^2 \frac{M_{N_1}}{M_{N_2}}+\left(y_{1\alpha} y_{\alpha 3}^{\dagger}\right)^2 \frac{M_{N_1}}{M_{N_3}} \right| \\
    &=-\frac{3\delta_{\text{eff}}}{16 \pi \Tilde{y}_{1i}\Tilde{y}_{i 1}^{\dagger} }  \left|\left(\Tilde{y}_{1j} \Tilde{y}_{j2}^{\dagger}\right)^2 \frac{M_{N_1}}{M_{N_2}}+\left(\Tilde{y}_{1k} \Tilde{y}_{k3}^{\dagger}\right)^2 \frac{M_{N_1}}{M_{N_3}} \right| \\
    &\simeq -\frac{3\delta_{\text{eff}}}{16\pi}  \Tilde{y}_{1i}\Tilde{y}_{i 1}^{\dagger}  \frac{M_{N_1}}{M_{N_2}} \\
    &\simeq \frac{3}{8\pi} \frac{M_{N_1}^2 m_3}{M_{N_2}v^2} \delta_{\text{eff}},
\end{split}
\end{equation}
In the first line in (\ref{eq:microscopic_lepton_asymmetry}), we used \(M_{N_{\gamma\neq 1}} \gg M_{N_1}\) for the approximation. In the second line, we have defined the effective CP phase in the Yukawa coupling matrix:

\begin{equation}
    \delta_{\text{eff}} \equiv \frac{\text{Im}\left[\left(y_{1\alpha} y_{\alpha 2}^{\dagger}\right)^2 \frac{M_{N_1}}{M_{N_2}}+\left(y_{1\alpha} y_{\alpha 3}^{\dagger}\right)^2 \frac{M_{N_1}}{M_{N_3}}\right]}{\left|\left(y_{1\alpha} y_{\alpha 2}^{\dagger}\right)^2 \frac{M_{N_1}}{M_{N_2}}+\left(y_{1\alpha} y_{\alpha 3}^{\dagger}\right)^2 \frac{M_{N_1}}{M_{N_3}} \right|} .
\end{equation}
Note that this quantity is less than unity by definition, \(\delta_{\text{eff}} \leq 1\). In the third line, we used the relation \(y_{1\alpha} y^\dagger_{\alpha \gamma} = \Tilde{y}_{1i} U^\dagger_{i \alpha} U_{\alpha j} \Tilde{y}^\dagger_{j \gamma} = \Tilde{y}_{1i} \delta_{ij} \Tilde{y}^\dagger_{j\gamma} = \Tilde{y}_{1i}\Tilde{y}^\dagger_{i\gamma}\). In the fourth line, we used the assumption that Yukawa coupling matrix is approximately proportional to the unit lower-right triangular matrix and \(M_{N_2} \ll M_{N_3}\), as mentioned in \ref{subsec:decay_of_majorana_neutrino}. \par

The expression for \(\delta\) \eqref{eq:microscopic_lepton_asymmetry} is different from the well-known expression (see  Eq.~(21) in \cite{gorbunov2011scalaron}). However, the expression for \(\Gamma_{N\to R}\) (see Eq.~(\ref{eq:decay_rate_NtoR_caseA})) and \(\delta\) in \cite{gorbunov2011scalaron} are not compatible. This is because the former assumes that \(m_3 \propto |y_{13}|^2/M_1\) so that the Yukawa coupling parameters in \(\Gamma_{N_1 \to R}\) can be eliminated, but the latter assumes, that \(m_3\propto |y_{33}|^2/M_3\) so that the \(M_3\) in \(\delta\) can be eliminated. That is, the largest contribution to \(m_3\) comes from \(M_{N_1}\) term in the former case, but from \(M_{N_3}\) term in the latter case. \par

The matching condition for the observed baryon-to-entropy ratio \((n_B/s) |_0\simeq 0.87\times 10^{-10}\) imposes restriction on the mass parameters.
The lepton asymmetry is generated through the non-thermal leptogenesis \cite{asaka1999leptogenesis,asaka2002non} because of the low reheating temperature and the large Majorana fermion mass, which realizes sizable gravitational particle production. Hence, there is no wash-out effect, and the Boltzmann equation for the lepton
number is simply given as
\begin{equation}
    \label{eq:lepton_asymmetry}
    \frac{dn_{L} a^3}{dt} = \delta \cdot \Gamma_{N\to R} n a^3.
\end{equation}
In terms of a rescaled variable: \(\bar{n}_{L} \equiv n_{L} \bar{a}^3 M^{-3}\), we find

\begin{equation}
    \frac{d\bar{n}_{L}}{d\bar{a}} = \delta\cdot \frac{\sqrt{3}M_{\text{G}} \Gamma_{N_1\to R} n \bar{a}}{M^2 \sqrt{r+f \bar{a} +n (M_{N_1}/M) \bar{a}}}.
    \label{eq:boltzman_lepton_asymmetry}
\end{equation}

The generated lepton asymmetry \(n_{L}\) can be calculated by integrating (\ref{eq:boltzman_lepton_asymmetry}). 
The lepton asymmetry is mostly generated around the end of the reheating. Once the relativistic particles are thermalized, \(n_L/s\) is given by

\begin{equation}
    \frac{n_L}{s} =\left(\frac{3^5\cdot 5}{2^7 \pi^2  g_r \rho_{R}^3}\right)^{\frac{1}{4}}\cdot  \frac{\bar{n}_L M^3}{\bar{a}^3}. 
\end{equation}
The lepton asymmetry is eventually converted to the baryon asymmetry by the electroweak sphaleron process to reach at~\cite{Harvey:1990qw}

\begin{equation}
    \left.\frac{n_B}{s}\right|_{0} \simeq -\left.\frac{28}{79} \frac{n_L}{s}\right|_\mathrm{reheating}.
\end{equation}

\begin{figure}[htbp]
    \centering
    \includegraphics[width=11cm,clip]{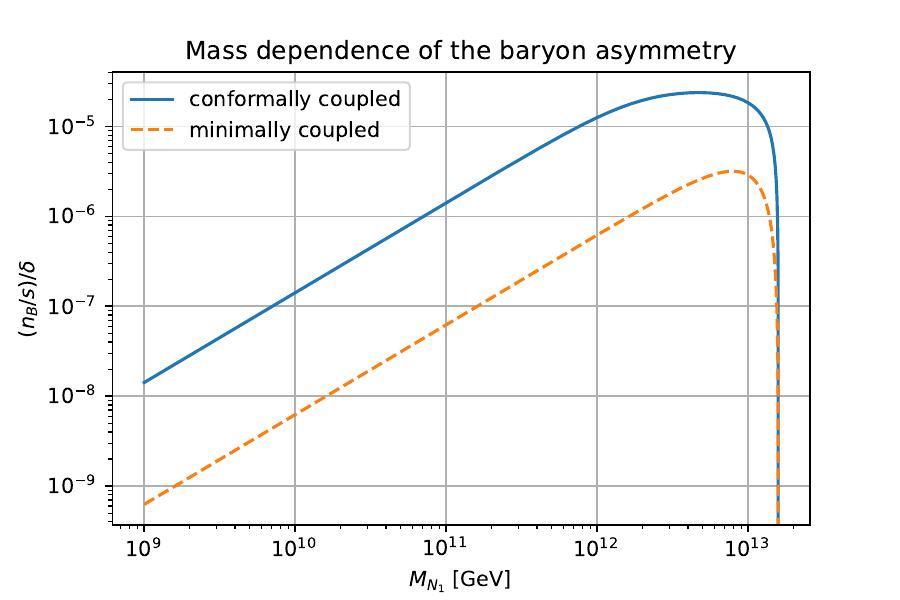}
    \caption{The baryon-to-entropy ratio normalized by \(\delta\), \(\left(n_B/ s\right)/ \delta \) is plotted with respect to the Majorana mass \(M_{N_1}\). The blue solid line and the orange dashed line represent the mass dependence of \(n_B/s) /\delta\) in the conformally coupled case and in the minimally coupled case. }
    \label{fig:baryon_asymmetry}
\end{figure}

We calculated the mass dependence of the resultant baryon-to-entropy ratio normalized by $\delta$, \((n_B/s)/ \delta \),   in the conformally coupled case and in the minimally coupled case in Fig.~\ref{fig:baryon_asymmetry}. We can see that the produced baryon asymmetry is larger in the case
of the conformally coupled case, with the same mass dependence, which is due to the enhancement of the branching ratio of \(\Gamma_{\phi \to N_1}\).\par

Next, we discuss the value of the mass \(M_{N_1}\) at which \((n_B/s)/ \delta \) takes the maximum. In the minimally coupled case, the entropy \(s\) is governed by the Higgs particles, but not by the decay products by the Majorana neutrino, and hence it is less affected by the change in mass \(M_{N_1}\). 
As a result, \((n_B/s)/ \delta \) is maximal near the mass at which \(\Gamma_{\phi\to N_1} \) (see (\ref{eq:decay_rates_majorna})) is 
maximal (\(M_{N_1}\simeq \SI{e13}{GeV}\)), when $n_B$ becomes maximum. On the other hand, in the conformally coupled case, the entropy \(s\) is governed by the decay products of the Majorana neutrinos, so is affected by the change in the mass \(M_{N_1}\). Thus, the mass \(M_{N_1}\) at which \((n_B/s)/ \delta \) is maximal deviates from \(M_{N_1}\simeq \SI{e13}{GeV}\) since the entropy density is also larger for heavier right-handed neutrino.\par

\begin{figure}[htbp]
    \centering
    \includegraphics[width=11cm,clip]{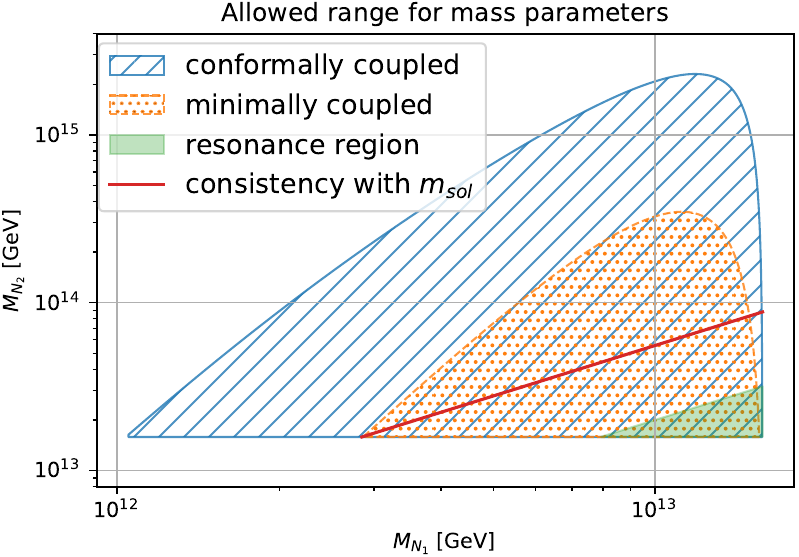}
    \caption{The allowed region for the mass parameters are shown. The blue ray-shaded region and the orange dot-shaded region represent the conformally coupled case and the minimally coupled case, respectively. In the green shaded region, the assumption on the hierarchy between $M_{N_1}$ and $M_{N_2}$ we made is broken and hence the result is not reliable there. The mass configuration must be on the red solid line in order to be consistent with the solar neutrino oscillation experiment. }
    \label{fig:allowd_mass}
\end{figure}
The different behavior in Fig.~\ref{fig:baryon_asymmetry} means that there is a difference in the constraint in the parameter we have. We show the limitation on the mass parameters \(M_{N_1}\) and \(M_{N_2}\) in Fig.~\ref{fig:allowd_mass}. 
From Fig.~\ref{fig:baryon_asymmetry}, one can read off the required \(\delta\) to explain the baryon asymmetry of the Universe for each \(M_{N_1}\). The upper-bound of \(M_{N_2}\) can be obtained by the condition \(\delta_{\text{eff}}\leq 1\), which is shown in Fig.~\ref{fig:allowd_mass}. 
We can see that the allowed range for the mass parameters is larger in the conformally coupled case due to the enhanced efficiency in the leptogenesis. We simultaneously showed the region where our analysis is not reliable \(\left|M_{N_2}-M_{N_1}\right|<M_{N_1}\) because we used the assumption \(M_{N_{\gamma\neq 1}} \gg M_{N_1}\) in the first line of (\ref{eq:microscopic_lepton_asymmetry}). In general, the resonant leptogenesis \cite{Pilaftsis:2003gt} may occur in this region, which requires  further analysis. There, 
we also showed the mass configuration in which \(m_2\simeq m_{\text{sol}} =\SI{0.009}{eV}\) is realized. This condition constrains the mass of the second lightest right-handed Majorana neutrino \(M_{N_2}\). This constraint is strongly dependent on the assumption of case A on the neutrino Yukawa coupling matrix. In the next section, we will investigate a
case with a different assumption on the Yukawa coupling matrix.

\section{The effect of the Yukawa coupling matrix}\label{sec:Yukawa_matrix}

Throughout \S\ref{sec:model_description}~-~\S\ref{sec:Connection_to_the_observation}, we have postulated that the components of the Yukawa coupling matrix follow case A (see \S ~\ref{subsec:decay_of_majorana_neutrino}). In this section, we consider case B where the Yukawa coupling matrix has a hierarchical structure. Case B is based on the following two assumptions \cite{fukugita1986barygenesis,asaka1999leptogenesis}. 

\paragraph{\(\spadesuit\) On the right-handed Majorana neutrino sector \(M_\alpha\), \(y_{\alpha \beta}\) (case B):}

First, we postulate
\begin{equation}
    \frac{\left(\Tilde{y}_{1i} \Tilde{y}_{i 2}^\dagger \right)^2}{M_2} = \frac{\left(\Tilde{y}_{1i} \Tilde{y}_{i 3}^\dagger \right)^2}{M_3}
\end{equation}
so that the second term in the second line of Eq.~(\ref{eq:microscopic_lepton_asymmetry}) is comparable to the first term, which leads to the increment of the net lepton asymmetry \(\delta\). This means that there is a hierarchy in the Yukawa coupling parameters that compensates the hierarchy of mass \(M_2 \ll M_3\). Second, we assume
\begin{equation}
    \left|\Tilde{y}_{\alpha 3}\right| > \left|\Tilde{y}_{\alpha 2}\right| \gg \left|\Tilde{y}_{\alpha 1}\right| \ \ (\alpha = 1,3). 
\end{equation}
In this case, the expression for the heaviest active neutrino \(m_3\) (see Eq.~(\ref{eq:neutrino_experiment})) is changed as
\begin{equation}
    m_3\simeq -\frac{\left|\Tilde{y}_{33}\right|^2v^2}{2M_3}. 
\end{equation}
This is because the largest contribution to \(m_3\) in Eq.~(\ref{eq:active_mass_matrix}) becomes the \(\gamma=3\) channel due to the hierarchical structure. 
Note that we keep the assumption of the hierarchy of the Majorana masses, 
$M_1 \ll M_2 \ll M_3$ to use the expression of the first line in (\ref{eq:microscopic_lepton_asymmetry}).

Using this expression, we can derive the well-known expression for the lepton asymmetry~\(\delta\):
\begin{equation}
    \delta = \frac{3}{8 \pi} \delta_{\text{eff}} \frac{m_3 M_{N_1}}{v^2},
\end{equation}
which has only one parameter \(M_{N_1}\) contrary to case A (\ref{eq:microscopic_lepton_asymmetry}).
Furthermore, the decay rate of the Majorana neutrino \(N_1\) into relativistic particles (see Eq.~(\ref{eq:decay_rate_NtoR_caseA})) is also changed as

\begin{equation}
\label{eq:decay_rate_NtoR_hierarchy}
    \Gamma_{N\to R} \simeq \frac{M_{N_1}}{8\pi} \left|\tilde{y}_{13}\right|^2. 
\end{equation}
In this case, we cannot re-write the decay rate \(\Gamma_{N\to R}\) in terms of \(m_3\) contrary to case A. \par

Based on this, we solve the set of the Boltzmann equations and calculate the baryon asymmetry as we did in the previous sections. We show the results in various parameter configurations below. 
First, when \(\xi\) is minimally coupled, nothing peculiar happens because the reheating by the decay into the Higgs field dominates and the parameter dependence on \(M_{N_1}\) and \(\Tilde{y}_{13}\) are drowned out. Therefore, the reheating temperature and the spectral parameters, in this case, are the same as in the minimally coupled case in Sec.~\ref{sec:Connection_to_the_observation}, which is shown in Fig.~\ref{fig:Yukawa_spectral}. 

\begin{figure}[htbp]
    \centering
    \includegraphics[width=12cm,clip]{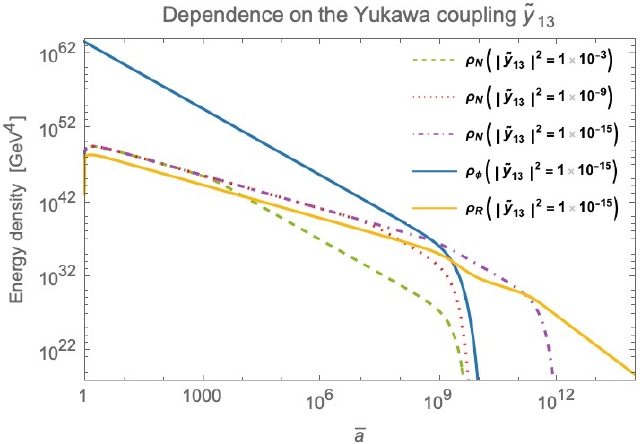}
    \caption{Results of numerical calculation of Boltzmann equations when the Yukawa coupling matrix has the hierarchical structure and the Higgs field is conformally coupled to gravity. The Majorana mass \(M_{N_1}\) is set to be \(\SI{1.0e13}{GeV}\). The green dashed line, the red dotted line and the purple dash-dot line represent the energy density of the Majorana neutirno \(N_1\) when the Yukawa coupling matrix component is  \(|\Tilde{y}_{13}|^2= \SI{1.0e-3}{}, \SI{1.0e-9}{} \text{and} \  \SI{1.0e-15}{}\), respectively. The blue solid line and the orange solid line represent the energy density of the inflaton and radiation, respectively. These lines are depicted for the case with \(|\Tilde{y}_{13}|^2= \SI{1.0e-15}{}\). }
    \label{fig:Yukawa_dependence}
\end{figure}

Second, when \(\xi\) is conformally coupled, the parameter dependence on the Majorana mass \(M_{N_1}\) and the Yukawa coupling parameter \(\Tilde{y}_{13}\) appears. We show the dependence of the energy densities on \(\Tilde{y}_{13}\) in Fig.~\ref{fig:Yukawa_dependence}. Here, the Majorana mass \(M_{N_1}\) is set to be \(\SI{1.0e13}{GeV}\).

This result shows the possibility of \(\rho_{N_1}\) surpassing \(\rho_{R}\) right after the end of the inflaton decay. This happens when the \(|\Tilde{y}_{13}|^2\lesssim \mathcal{O}(10^{-12})\). In this case, the reheating process is not completed by the inflaton decay, but by the subsequent decay of the right-handed Majorana neutrino. In consequence, the matter-dominated epoch lasts for a longer period and this results in the drastic change in the spectral parameters. The spectral parameters in this case are shown in Fig.~\ref{fig:Yukawa_spectral}. We plotted the region where the above \(N_1\) domination happen: \(\mathcal{O} (10^{-17}) \lesssim |\Tilde{y}_{13}|^2\lesssim \mathcal{O} (10^{-12})\). For \(|\Tilde{y}_{13}|^2 \gtrsim \mathcal{O} (10^{-11})\), this \(N_1\) domination does not occur and the resultant spectral parameters approach to the circle point in Fig.~\ref{fig:nsr}. \par

There is a restriction on the smallness of the Yukawa coupling parameter \(|\Tilde{y}_{13}|^2 \) when we consider realizing the observed baryon asymmetry of the universe: \((n_B/s)|_0 = \SI{0.87e-10}{}\). The lower-bound on \(|\Tilde{y}_{13}|^2\) can be obtained by integrating the Eq.~(\ref{eq:boltzman_lepton_asymmetry}) and using \(\delta_{\text{eff}} \leq 1\) as we did in Fig.~\ref{fig:allowd_mass}. The results are shown in Fig.~\ref{fig:baryon_asymmetry_YM_dependence} as the solid line. This argument leads to the condition: 
\begin{equation}
    |\Tilde{y}_{13}|^2 > \mathcal{O} (10^{-17}). 
\end{equation}

We have fixed the Majorana mass to be \(\SI{1.0e13}{GeV}\) above. For smaller masses, the production rate of the Majorana neutrinos becomes smaller (see Eq.~(\ref{eq:decay_rates_majorna})), so that the above matter dominance period by the Majorana neutrino does not happen for \(M_{N_1} \lesssim \mathcal{O} (\SI{e11}{GeV})\). The resultant spectral parameters are close to case A, which is shown in Fig.~\ref{fig:nsr}. In this case, the lower bound on the Yukawa coupling parameter \(|\Tilde{y}_{13}|^2\) becomes stricter as we decrease the mass \(M_{N_1}\). This is shown in Fig.~\ref{fig:baryon_asymmetry_YM_dependence}. If the mass drops to \(M_{N_1} \lesssim \SI{2.5e11}{GeV}\), any values for the Yukawa coupling parameter \(\Tilde{y}_{13}\) are forbidden, so we find the Majorana mass must satisfy
\begin{equation}
    M_{N_1} \gtrsim \SI{2.5e11}{GeV}
\end{equation}
to be consistent with the baryon asymmetry of the universe. 

\section{Conclusion and  Discussion}\label{sec:Conclusion_Discussion}

We have analyzed the \(R^2\) inflationary model with an extended matter sector containing right-handed Majorana neutrinos which are responsible for the  leptogenesis  of the baryon asymmetry in the Universe in order to make it a more realistic cosmological model. We conducted a comprehensive study of the reheating process by considering all of the matter fields. We  identified the relevant decay channels and solved the evolution equations of the system. 
There are several parameters that characterize this model: the non-minimal coupling \(\xi\), the mass of the Majorana fermion \(M_{N_1}\), and the Yukawa coupling matrix \(y_{\alpha \beta}\).

\begin{figure}[H]
    \centering
    \includegraphics[width=10cm,clip]{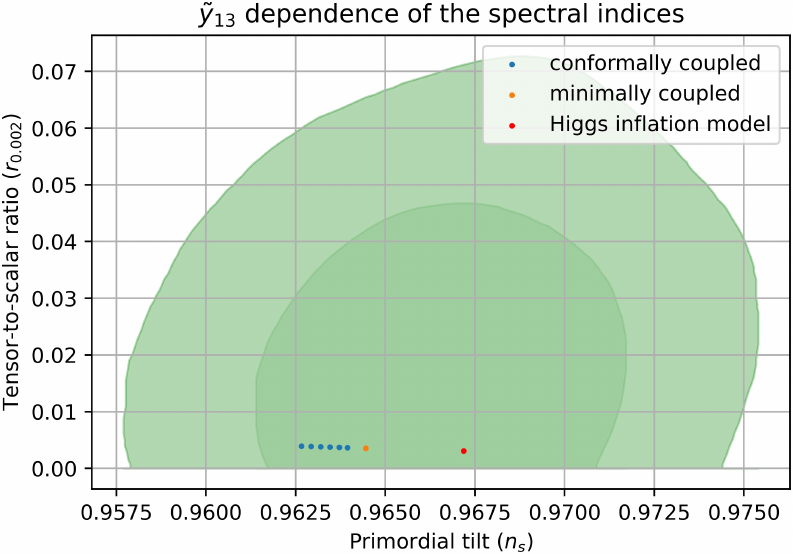}
    \caption{The spectral parameters for various values of Yukawa couplings are shown. The mass of the right-handed Majorana neutrino \(M_{N_1}\) is fixed to be \(\SI{1.0e13}{GeV}\). The blue plot and the orange plot represent the \(y_{31}\) dependence of the spectral parameters \(n_s,r\) in the conformally coupled case and minimally coupled case, respectively. The multiple blue plot represent the value when \(|\Tilde{y}_{13}|^2 = \SI{1.0e-17}{}, \ \SI{1.0e-16}{}, \ \SI{1.0e-15}{}, \ \SI{1.0e-14}{}, \ \SI{1.0e-13}{}, \ \SI{1.0e-12}{}\) from the left to the right. The value for \(\SI{1.0e-17}{}\) is not allowed in view of realizing the observed baryon asymmetry. The \(68 \%\) and \(95 \%\) CL region for the spectral parameters are simultaneously shown as same as in Fig.~\ref{fig:nsr}. }
    \label{fig:Yukawa_spectral}
\end{figure}

\begin{figure}[H]
    \centering
    \includegraphics[width=11cm,clip]{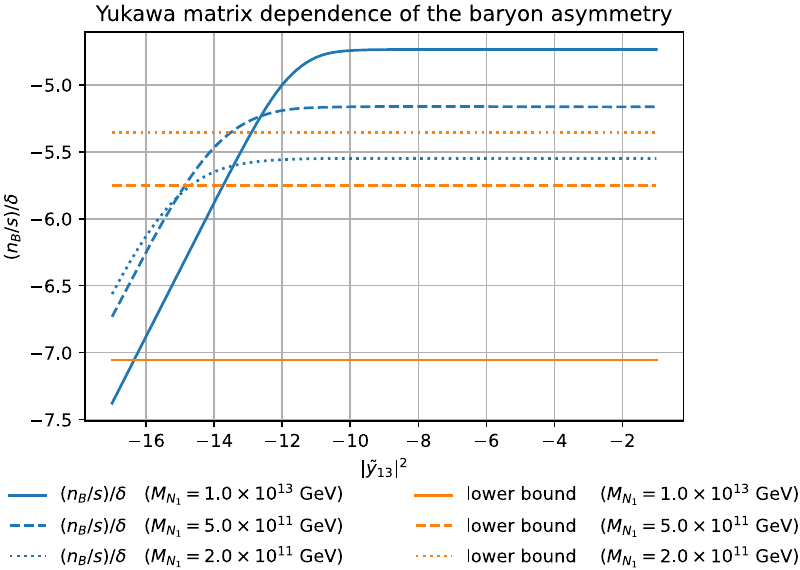}
    \caption{The baryon-to-entropy ratio \((n_B/s)/\delta\) is plotted with respect to the Yukawa coupling parameter \(|\Tilde{y}_{13}|^2\) for various Majorana masses \(M_{N_1}\). The blue lines and the orange lines represent the ratio \((n_B/s)/\delta\) and the lower bound on the ratio coming from \(\delta_{\text{eff}} \leq 1\). 
    In other words, the present baryon asymmetry of the Universe can be explained if the blue line exceeds the orange line for each choice of $M_{N_1}$. The solid lines, dashed lines and dotted lines represent the case when the Majorana mass is \(\SI{1.0e13}{GeV}\), \(\SI{5.0e11}{GeV}\) and \(\SI{2.0e11}{GeV}\), respectively. }
    \label{fig:baryon_asymmetry_YM_dependence}
\end{figure}

When the Higgs field is minimally coupled with $\xi=0$, reheating is entirely dominated by it and parameters of the neutrino sector affect only the details of leptogenesis.  When it is conformally coupled with $\xi=-1/6$, on the other hand, we observe nontrivial effects of the neutrino sector on reheating both for the cases A and B, which we highlight below.\par

When the Yukawa coupling matrix \(y_{\alpha \beta}\) has the structure of case A and the Higgs field is conformally coupled, the decay rate \(\Gamma_{\phi\to h}\) is much suppressed, thus decay into the right-handed Majorana neutrinos and into gauge bosons becomes relevant. Which decay is more relevant is determined by the Majorana mass \(M_{N_1}\), so the mass dependence appears. 
We have confirmed this behavior by examining the mass dependence of the reheating process and the observables: the spectral parameters of primordial perturbations and the baryon asymmetry. 
As we can see in Fig.~\ref{fig:nsr} and Fig.~\ref{fig:baryon_asymmetry}, the mass dependence appears strongly in the conformally coupled case and not so much in the minimally coupled case. Furthermore, we showed the limitation on the mass parameters by considering the consistency with the value of the present baryon-to-photon ratio estimated from observations. In this case, the Yukawa coupling matrix is chosen such that we have $m_3=m_\mathrm{atm}$ and no further freedom to change.
\par

On the contrary, in the case B in which the Yukawa coupling matrix has a hierarchical structure, 
the value of the Yukawa coupling matrix comes into play. 
In this case, we have three parameters which characterize the model: \(\xi\), \(M_{N_1}\), \(\Tilde{y}_{13}\). When the Higgs field is conformally coupled, the parameters in the neutrino sector influence the model. In particular, when the Majorana mass is large and the Yukawa coupling \(\Tilde{y}_{13}\) is small, the Majorana neutrinos can dominate the energy density of the universe even after the inflaton decay as is shown in Fig.~\ref{fig:Yukawa_dependence}. This leads to the longer matter-dominated era and thus results in the drastic change in the spectral parameters, which are shown in Fig.~\ref{fig:Yukawa_spectral}. If we decrease the Majorana mass \(M_{N_1}\), the dependence on the Yukawa coupling disappears and the reheating process is governed by the decay of the inflaton into gauge bosons. The constraints on the parameters are also obtained in this case, see Fig.~\ref{fig:baryon_asymmetry_YM_dependence}. \par

In the present study, we assumed the hierarchy between the Majorana neutrino mass and 
the scalaron mass, $M_1<M/2<M_2$. 
If we consider the case $M_1 \ll M_2<M/2$, another scenario can be realized 
where the reheating and baryogenesis is mainly governed by the $N_2$ decay
while $N_1$ subsequently generated can be a dark matter candidate, 
in a similar way considered in Ref.~\cite{Fujikura:2022udt}.

Also, we only considered the cases with minimal and conformal couplings of the Higgs field to gravity.
If we take $\xi>0$, it has been found that the Higgs field is tachyonic along
the $R^2$ inflationary trajectory and the system enters the regime of the mixed Higgs-$R^2$ model~\cite{ema2017higgs,Wang:2017fuy,he2018inflation}. 
Once we introduce massive right-handed neutrinos into the mixed Higgs-$R^2$ model, we may have a decay channel from Higgs to right-handed neutrinos for sufficiently large $\xi$
through non-perturbative Higgs production~\cite{Bezrukov:2019ylq,He:2020ivk}.
The detailed study of the effect of right-handed neutrinos 
on reheating and baryogenesis in this case is left for future study.\par

\acknowledgments

We thank A. Kamada, Y. Watanabe and K. Mukaida for useful discussion. The work of H.~J. is supported by the Forefront Physics and Mathematics Program to Drive Transformation (FoPM). K.\,K.\, is supported by JSPS KAKENHI Grant-in-Aid for Challenging Research (Exploratory) JP23K17687. 
A.A.S. was partially supported by the RSF grant 21-12-00130. J.Y. is supported by JSPS KAKENHI Grant (S) No.\ 20H05639.

\bibliographystyle{utphys.bst}
\bibliography{paper1}

\providecommand{\href}[2]{#2}\begingroup\raggedright\begin{thebibliography}{10}

\bibitem{starobinsky1980new}
A.~A. Starobinsky, ``A new type of isotropic cosmological models without
  singularity,'' {\em Physics Letters B} {\bfseries 91} no.~1, (1980) 99--102.

\bibitem{sato1981first}
K.~Sato, ``First-order phase transition of a vacuum and the expansion of the
  universe,'' {\em Monthly Notices of the Royal Astronomical Society}
  {\bfseries 195} no.~3, (1981) 467--479.

\bibitem{guth1981inflationary}
A.~H. Guth, ``Inflationary universe: A possible solution to the horizon and
  flatness problems,'' {\em Physical Review D} {\bfseries 23} no.~2, (1981)
  347.

\bibitem{linde1982new}
A.~D. Linde, ``A new inflationary universe scenario: a possible solution of the
  horizon, flatness, homogeneity, isotropy and primordial monopole problems,''
  {\em Physics Letters B} {\bfseries 108} no.~6, (1982) 389--393.

\bibitem{albrecht1982cosmology}
A.~Albrecht and P.~J. Steinhardt, ``Cosmology for grand unified theories with
  radiatively induced symmetry breaking,'' {\em Physical Review Letters}
  {\bfseries 48} no.~17, (1982) 1220.

\bibitem{sato2015inflationary}
K.~Sato and J.~Yokoyama, ``Inflationary cosmology: First 30+ years,'' {\em
  International Journal of Modern Physics D} {\bfseries 24} no.~11, (2015)
  1530025.

\bibitem{Starobinsky:1981vz}
A.~A. Starobinsky, ``{Nonsingular Model of the Universe with the Quantum
  Gravitational De Sitter Stage and its Observational Consequences},'' in {\em
  {Proc. of the Second Seminar ``Quantum Theory of Gravity'', Moscow Oct. 1981,
  INR Press, Moscow (1982)}}, pp.~58--72.

\bibitem{Starobinsky:1983zz}
A.~A. Starobinsky, ``{The Perturbation Spectrum Evolving from a Nonsingular
  Initially De-Sitter Cosmology and the Microwave Background Anisotropy},''
  {\em Sov. Astron. Lett.} {\bfseries 9} (1983) 302--304.

\bibitem{Vilenkin:1985md}
A.~Vilenkin, ``{Classical and Quantum Cosmology of the Starobinsky Inflationary
  Model},'' \href{http://dx.doi.org/10.1103/PhysRevD.32.2511}{{\em Phys. Rev.
  D} {\bfseries 32} (1985) 2511}.

\bibitem{Mijic:1986iv}
M.~B. Mijic, M.~S. Morris, and W.-M. Suen, ``{The R**2 Cosmology: Inflation
  Without a Phase Transition},''
  \href{http://dx.doi.org/10.1103/PhysRevD.34.2934}{{\em Phys. Rev. D}
  {\bfseries 34} (1986) 2934}.

\bibitem{Maeda:1987xf}
K.-i. Maeda, ``{Inflation as a Transient Attractor in R**2 Cosmology},''
  \href{http://dx.doi.org/10.1103/PhysRevD.37.858}{{\em Phys. Rev. D}
  {\bfseries 37} (1988) 858}.

\bibitem{Zeldovich:1971mw}
Y.~B. Zeldovich and A.~A. Starobinsky, ``{Particle production and vacuum
  polarization in an anisotropic gravitational field},'' {\em Sov. Phys. JETP}
  {\bfseries 34} (1972) 1159--1166.

\bibitem{Parker:1974qw}
L.~Parker and S.~A. Fulling, ``{Adiabatic regularization of the energy momentum
  tensor of a quantized field in homogeneous spaces},''
  \href{http://dx.doi.org/10.1103/PhysRevD.9.341}{{\em Phys. Rev. D} {\bfseries
  9} (1974) 341--354}.

\bibitem{birrell1984quantum}
N.~D. Birrell and P.~C.~W. Davies, {\em Quantum fields in curved space}.
\newblock Cambridge University press, 1984.

\bibitem{Nariai:1971sv}
H.~Nariai and K.~Tomita, ``{On the removal of initial singularity in a big-bang
  universe in terms of a renormalized theory of gravitation. 2. criteria for
  obtaining a physically reasonable model},''
  \href{http://dx.doi.org/10.1143/PTP.46.776}{{\em Prog. Theor. Phys.}
  {\bfseries 46} (1971) 776--786}.

\bibitem{faulkner2007constraining}
T.~Faulkner, M.~Tegmark, E.~F. Bunn, and Y.~Mao, ``Constraining f (r) gravity
  as a scalar-tensor theory,'' {\em Physical Review D} {\bfseries 76} no.~6,
  (2007) 063505.

\bibitem{netto2016stable}
T.~d.~P. Netto, A.~M. Pelinson, I.~L. Shapiro, and A.~A. Starobinsky, ``From
  stable to unstable anomaly-induced inflation,'' {\em The European Physical
  Journal C} {\bfseries 76} no.~10, (2016) 544.

\bibitem{cervantes1995induced}
J.~L. Cervantes-Cota and H.~Dehnen, ``Induced gravity inflation in the standard
  model of particle physics,'' {\em Nuclear Physics B} {\bfseries 442} no.~1-2,
  (1995) 391--409.

\bibitem{bezrukov2008standard}
F.~Bezrukov and M.~Shaposhnikov, ``The standard model higgs boson as the
  inflaton,'' {\em Physics Letters B} {\bfseries 659} no.~3, (2008) 703--706.

\bibitem{Barvinsky:2008ia}
A.~O. Barvinsky, A.~Y. Kamenshchik, and A.~A. Starobinsky, ``{Inflation
  scenario via the Standard Model Higgs boson and LHC},''
  \href{http://dx.doi.org/10.1088/1475-7516/2008/11/021}{{\em JCAP} {\bfseries
  11} (2008) 021}, \href{http://arxiv.org/abs/0809.2104}{{\ttfamily
  arXiv:0809.2104 [hep-ph]}}.

\bibitem{bezrukov2009initial}
F.~Bezrukov, D.~Gorbunov, and M.~Shaposhnikov, ``On initial conditions for the
  hot big bang,'' {\em Journal of Cosmology and Astroparticle Physics}
  {\bfseries 2009} no.~06, (2009) 029.

\bibitem{Barvinsky:2009fy}
A.~O. Barvinsky, A.~Y. Kamenshchik, C.~Kiefer, A.~A. Starobinsky, and
  C.~Steinwachs, ``{Asymptotic freedom in inflationary cosmology with a
  non-minimally coupled Higgs field},''
  \href{http://dx.doi.org/10.1088/1475-7516/2009/12/003}{{\em JCAP} {\bfseries
  12} (2009) 003}, \href{http://arxiv.org/abs/0904.1698}{{\ttfamily
  arXiv:0904.1698 [hep-ph]}}.

\bibitem{garcia2009preheating}
J.~Garcia-Bellido, D.~G. Figueroa, and J.~Rubio, ``Preheating in the standard
  model with the higgs inflaton coupled to gravity,'' {\em Physical Review D}
  {\bfseries 79} no.~6, (2009) 063531.

\bibitem{ema2017violent}
Y.~Ema, R.~Jinno, K.~Mukaida, and K.~Nakayama, ``Violent preheating in
  inflation with nonminimal coupling,'' {\em Journal of Cosmology and
  Astroparticle Physics} {\bfseries 2017} no.~02, (2017) 045.

\bibitem{decross2018preheating}
M.~P. DeCross, D.~I. Kaiser, A.~Prabhu, C.~Prescod-Weinstein, and E.~I.
  Sfakianakis, ``Preheating after multifield inflation with nonminimal
  couplings. i. covariant formalism and attractor behavior,'' {\em Physical
  Review D} {\bfseries 97} no.~2, (2018) 023526.

\bibitem{Planck:2018jri}
{\bfseries Planck} Collaboration, Y.~Akrami {\em et~al.}, ``{Planck 2018
  results. X. Constraints on inflation},''
  \href{http://dx.doi.org/10.1051/0004-6361/201833887}{{\em Astron. Astrophys.}
  {\bfseries 641} (2020) A10},
  \href{http://arxiv.org/abs/1807.06211}{{\ttfamily arXiv:1807.06211
  [astro-ph.CO]}}.

\bibitem{He:2018gyf}
M.~He, A.~A. Starobinsky, and J.~Yokoyama, ``{Inflation in the mixed
  Higgs-$R^2$ model},''
  \href{http://dx.doi.org/10.1088/1475-7516/2018/05/064}{{\em JCAP} {\bfseries
  05} (2018) 064}, \href{http://arxiv.org/abs/1804.00409}{{\ttfamily
  arXiv:1804.00409 [astro-ph.CO]}}.

\bibitem{gorbunov2011scalaron}
D.~S. Gorbunov and A.~G. Panin, ``Scalaron the mighty: producing dark matter
  and baryon asymmetry at reheating,'' {\em Physics Letters B} {\bfseries 700}
  no.~3-4, (2011) 157--162.

\bibitem{gorbunov2012free}
D.~S. Gorbunov and A.~G. Panin, ``Free scalar dark matter candidates in
  r2-inflation: the light, the heavy and the superheavy,'' {\em Physics Letters
  B} {\bfseries 718} no.~1, (2012) 15--20.

\bibitem{CMB-S4:2016ple}
{\bfseries CMB-S4} Collaboration, K.~N. Abazajian {\em et~al.}, ``{CMB-S4
  Science Book, First Edition},''  (10, 2016) ,
  \href{http://arxiv.org/abs/1610.02743}{{\ttfamily arXiv:1610.02743
  [astro-ph.CO]}}.

\bibitem{Abazajian:2019eic}
K.~Abazajian {\em et~al.}, ``{CMB-S4 Science Case, Reference Design, and
  Project Plan},''  (7, 2019) ,
  \href{http://arxiv.org/abs/1907.04473}{{\ttfamily arXiv:1907.04473
  [astro-ph.IM]}}.

\bibitem{LiteBIRD:2022cnt}
{\bfseries LiteBIRD} Collaboration, E.~Allys {\em et~al.}, ``{Probing Cosmic
  Inflation with the LiteBIRD Cosmic Microwave Background Polarization
  Survey},'' \href{http://dx.doi.org/10.1093/ptep/ptac150}{{\em PTEP}
  {\bfseries 2023} no.~4, (2023) 042F01},
  \href{http://arxiv.org/abs/2202.02773}{{\ttfamily arXiv:2202.02773
  [astro-ph.IM]}}.

\bibitem{gorbunov2013r2}
D.~Gorbunov and A.~Tokareva, ``R2-inflation with conformal sm higgs field,''
  {\em Journal of Cosmology and Astroparticle Physics} {\bfseries 2013} no.~12,
  (2013) 021.

\bibitem{fukugita1986barygenesis}
M.~Fukugita and T.~Yanagida, ``Barygenesis without grand unification,'' {\em
  Physics Letters B} {\bfseries 174} no.~1, (1986) 45--47.

\bibitem{asaka1999leptogenesis}
T.~Asaka, K.~Hamaguchi, M.~Kawasaki, and T.~Yanagida, ``Leptogenesis in
  inflaton decay,'' {\em Physics Letters B} {\bfseries 464} no.~1-2, (1999)
  12--18.

\bibitem{maeda1989towards}
K.-i. Maeda, ``Towards the einstein-hilbert action via conformal
  transformation,'' {\em Physical Review D} {\bfseries 39} no.~10, (1989) 3159.

\bibitem{watanabe2011rate}
Y.~Watanabe, ``Rate of gravitational inflaton decay via gauge trace anomaly,''
  {\em Physical Review D} {\bfseries 83} no.~4, (2011) 043511.

\bibitem{morozov1986anomalies}
A.~Morozov, ``Anomalies in gauge theories,'' {\em Soviet Physics Uspekhi}
  {\bfseries 29} no.~11, (1986) 993.

\bibitem{Kamada:2019pmx}
A.~Kamada, ``{On Scalaron Decay via the Trace of Energy-Momentum Tensor},''
  \href{http://dx.doi.org/10.1007/JHEP07(2019)172}{{\em JHEP} {\bfseries 07}
  (2019) 172}, \href{http://arxiv.org/abs/1902.05209}{{\ttfamily
  arXiv:1902.05209 [hep-ph]}}.

\bibitem{Kamada:2019hpp}
A.~Kamada and T.~Kuwahara, ``{Lessons from $T^{\mu}_{~ \mu}$ on inflation
  models: Two-loop renormalization of $\eta$ in the scalar QED},''
  \href{http://dx.doi.org/10.1103/PhysRevD.103.116001}{{\em Phys. Rev. D}
  {\bfseries 103} no.~11, (2021) 116001},
  \href{http://arxiv.org/abs/1909.04229}{{\ttfamily arXiv:1909.04229
  [hep-ph]}}.

\bibitem{Kamada:2019euz}
A.~Kamada and T.~Kuwahara, ``{Lessons from $T^{\mu}_{~ \mu}$ on inflation
  models: Two-scalar theory and Yukawa theory},''
  \href{http://dx.doi.org/10.1103/PhysRevD.101.096012}{{\em Phys. Rev. D}
  {\bfseries 101} no.~9, (2020) 096012},
  \href{http://arxiv.org/abs/1909.04228}{{\ttfamily arXiv:1909.04228
  [hep-ph]}}.

\bibitem{dreiner2010two}
H.~K. Dreiner, H.~E. Haber, and S.~P. Martin, ``Two-component spinor techniques
  and feynman rules for quantum field theory and supersymmetry,'' {\em Physics
  Reports} {\bfseries 494} no.~1-2, (2010) 1--196.

\bibitem{case1957reformulation}
K.~M. Case, ``Reformulation of the majorana theory of the neutrino,'' {\em
  Physical Review} {\bfseries 107} no.~1, (1957) 307.

\bibitem{pal2011dirac}
P.~B. Pal, ``Dirac, majorana, and weyl fermions,'' {\em American Journal of
  Physics} {\bfseries 79} no.~5, (2011) 485--498.

\bibitem{Starobinsky:1981zc}
A.~A. Starobinsky, ``{Evolution of small perturbations of isotropic
  cosmological models with one-loop quantum gravitational corrections},'' {\em
  JETP Lett.} {\bfseries 34} (1981) 438--441.

\bibitem{Ema:2016hlw}
Y.~Ema, R.~Jinno, K.~Mukaida, and K.~Nakayama, ``{Gravitational particle
  production in oscillating backgrounds and its cosmological implications},''
  {\em Phys. Rev. D} {\bfseries 94} (2016) 063517,
  \href{http://arxiv.org/abs/1604.08898}{{\ttfamily 1604.08898}}.

\bibitem{Buttazzo:2013uya}
D.~Buttazzo, G.~Degrassi, P.~P. Giardino, G.~F. Giudice, F.~Sala, A.~Salvio,
  and A.~Strumia, ``{Investigating the near-criticality of the Higgs boson},''
  \href{http://dx.doi.org/10.1007/JHEP12(2013)089}{{\em JHEP} {\bfseries 12}
  (2013) 089}, \href{http://arxiv.org/abs/1307.3536}{{\ttfamily arXiv:1307.3536
  [hep-ph]}}.

\bibitem{Super-Kamiokande:1998kpq}
{\bfseries Super-Kamiokande} Collaboration, Y.~Fukuda {\em et~al.}, ``{Evidence
  for oscillation of atmospheric neutrinos},''
  \href{http://dx.doi.org/10.1103/PhysRevLett.81.1562}{{\em Phys. Rev. Lett.}
  {\bfseries 81} (1998) 1562--1567},
  \href{http://arxiv.org/abs/hep-ex/9807003}{{\ttfamily arXiv:hep-ex/9807003}}.

\bibitem{T2K:2014ghj}
{\bfseries T2K} Collaboration, K.~Abe {\em et~al.}, ``{Precise Measurement of
  the Neutrino Mixing Parameter $\theta_{23}$ from Muon Neutrino Disappearance
  in an Off-Axis Beam},''
  \href{http://dx.doi.org/10.1103/PhysRevLett.112.181801}{{\em Phys. Rev.
  Lett.} {\bfseries 112} no.~18, (2014) 181801},
  \href{http://arxiv.org/abs/1403.1532}{{\ttfamily arXiv:1403.1532 [hep-ex]}}.

\bibitem{gonzalez2012global}
M.~Gonzalez-Garcia, M.~Maltoni, J.~Salvado, and T.~Schwetz, ``Global fit to
  three neutrino mixing: critical look at present precision,'' {\em Journal of
  High Energy Physics} {\bfseries 2012} no.~12, (2012) 1--24.

\bibitem{chung1999production}
D.~J.~H. Chung, E.~W. Kolb, and A.~Riotto, ``Production of massive particles
  during reheating,'' {\em Physical Review D} {\bfseries 60} no.~6, (1999)
  063504.

\bibitem{Kolb:1990vq}
E.~W. Kolb and M.~S. Turner,
  \href{http://dx.doi.org/10.1201/9780429492860}{{\em {The Early Universe}}},
  vol.~69.
\newblock Nature Publishing Group UK London, 1990.

\bibitem{davidson2000thermalisation}
S.~Davidson and S.~Sarkar, ``Thermalisation after inflation,'' {\em Journal of
  High Energy Physics} {\bfseries 2000} no.~11, (2000) 012.

\bibitem{allahverdi2010reheating}
R.~Allahverdi, R.~Brandenberger, F.-Y. Cyr-Racine, and A.~Mazumdar, ``Reheating
  in inflationary cosmology: theory and applications,'' {\em Annual Review of
  Nuclear and Particle Science} {\bfseries 60} (2010) 27--51.

\bibitem{Harigaya:2013vwa}
K.~Harigaya and K.~Mukaida, ``{Thermalization after/during Reheating},''
  \href{http://dx.doi.org/10.1007/JHEP05(2014)006}{{\em JHEP} {\bfseries 05}
  (2014) 006}, \href{http://arxiv.org/abs/1312.3097}{{\ttfamily arXiv:1312.3097
  [hep-ph]}}.

\bibitem{Mukaida:2015ria}
K.~Mukaida and M.~Yamada, ``{Thermalization Process after Inflation and
  Effective Potential of Scalar Field},''
  \href{http://dx.doi.org/10.1088/1475-7516/2016/02/003}{{\em JCAP} {\bfseries
  02} (2016) 003}, \href{http://arxiv.org/abs/1506.07661}{{\ttfamily
  arXiv:1506.07661 [hep-ph]}}.

\bibitem{Mukaida:2022bbo}
K.~Mukaida and M.~Yamada, ``{Cascades of high-energy SM particles in the
  primordial thermal plasma},''
  \href{http://dx.doi.org/10.1007/JHEP10(2022)116}{{\em JHEP} {\bfseries 10}
  (2022) 116}, \href{http://arxiv.org/abs/2208.11708}{{\ttfamily
  arXiv:2208.11708 [hep-ph]}}.

\bibitem{baumann2022cosmology}
D.~Baumann, {\em Cosmology}.
\newblock Cambridge University Press, 2022.

\bibitem{Manton:1983nd}
N.~S. Manton, ``{Topology in the Weinberg-Salam Theory},''
  \href{http://dx.doi.org/10.1103/PhysRevD.28.2019}{{\em Phys. Rev. D}
  {\bfseries 28} (1983) 2019}.

\bibitem{klinkhamer1984saddle}
F.~R. Klinkhamer and N.~S. Manton, ``A saddle-point solution in the
  weinberg-salam theory,'' {\em Physical Review D} {\bfseries 30} no.~10,
  (1984) 2212.

\bibitem{arnold1987sphalerons}
P.~Arnold and L.~McLerran, ``Sphalerons, small fluctuations, and baryon-number
  violation in electroweak theory,'' {\em Physical Review D} {\bfseries 36}
  no.~2, (1987) 581.

\bibitem{Kuzmin:1985mm}
V.~A. Kuzmin, V.~A. Rubakov, and M.~E. Shaposhnikov, ``{On the Anomalous
  Electroweak Baryon Number Nonconservation in the Early Universe},''
  \href{http://dx.doi.org/10.1016/0370-2693(85)91028-7}{{\em Phys. Lett. B}
  {\bfseries 155} (1985) 36}.

\bibitem{covi1996cp}
L.~Covi, E.~Roulet, and F.~Vissani, ``Cp violating decays in leptogenesis
  scenarios,'' {\em Physics Letters B} {\bfseries 384} no.~1-4, (1996)
  169--174.

\bibitem{asaka2002non}
T.~Asaka, H.~B. Nielsen, and Y.~Takanishi, ``Non-thermal leptogenesis from the
  heavier majorana neutrinos,'' {\em Nuclear Physics B} {\bfseries 647}
  no.~1-2, (2002) 252--274.

\bibitem{Harvey:1990qw}
J.~A. Harvey and M.~S. Turner, ``{Cosmological baryon and lepton number in the
  presence of electroweak fermion number violation},''
  \href{http://dx.doi.org/10.1103/PhysRevD.42.3344}{{\em Phys. Rev. D}
  {\bfseries 42} (1990) 3344--3349}.

\bibitem{Pilaftsis:2003gt}
A.~Pilaftsis and T.~E.~J. Underwood, ``{Resonant leptogenesis},''
  \href{http://dx.doi.org/10.1016/j.nuclphysb.2004.05.029}{{\em Nucl. Phys. B}
  {\bfseries 692} (2004) 303--345},
  \href{http://arxiv.org/abs/hep-ph/0309342}{{\ttfamily arXiv:hep-ph/0309342}}.

\bibitem{Fujikura:2022udt}
K.~Fujikura, S.~Hashiba, and J.~Yokoyama, ``{Generation of neutrino dark
  matter, baryon asymmetry, and radiation after quintessential inflation},''
  \href{http://dx.doi.org/10.1103/PhysRevD.107.063537}{{\em Phys. Rev. D}
  {\bfseries 107} no.~6, (2023) 063537},
  \href{http://arxiv.org/abs/2210.05214}{{\ttfamily arXiv:2210.05214
  [hep-ph]}}.

\bibitem{ema2017higgs}
Y.~Ema, ``Higgs scalaron mixed inflation,'' {\em Physics Letters B} {\bfseries
  770} (2017) 403--411.

\bibitem{Wang:2017fuy}
Y.-C. Wang and T.~Wang, ``{Primordial perturbations generated by Higgs field
  and $R^2$ operator},''
  \href{http://dx.doi.org/10.1103/PhysRevD.96.123506}{{\em Phys. Rev. D}
  {\bfseries 96} no.~12, (2017) 123506},
  \href{http://arxiv.org/abs/1701.06636}{{\ttfamily arXiv:1701.06636 [gr-qc]}}.

\bibitem{he2018inflation}
M.~He, A.~A. Starobinsky, and J.~Yokoyama, ``Inflation in the mixed higgs-r2
  model,'' {\em Journal of Cosmology and Astroparticle Physics} {\bfseries
  2018} no.~05, (2018) 064.

\bibitem{Bezrukov:2019ylq}
F.~Bezrukov, D.~Gorbunov, C.~Shepherd, and A.~Tokareva, ``{Some like it hot:
  $R^2$ heals Higgs inflation, but does not cool it},''
  \href{http://dx.doi.org/10.1016/j.physletb.2019.06.064}{{\em Phys. Lett. B}
  {\bfseries 795} (2019) 657--665},
  \href{http://arxiv.org/abs/1904.04737}{{\ttfamily arXiv:1904.04737
  [hep-ph]}}.

\bibitem{He:2020ivk}
M.~He, R.~Jinno, K.~Kamada, A.~A. Starobinsky, and J.~Yokoyama, ``{Occurrence
  of tachyonic preheating in the mixed Higgs-R$^2$ model},''
  \href{http://dx.doi.org/10.1088/1475-7516/2021/01/066}{{\em JCAP} {\bfseries
  01} (2021) 066}, \href{http://arxiv.org/abs/2007.10369}{{\ttfamily
  arXiv:2007.10369 [hep-ph]}}.

\end{thebibliography}\endgroup

\end{document}